%% file: quantum_version.tex
\documentclass[a4paper,onecolumn, 11pt,accepted=2026-05-25]{quantumarticle}
\pdfoutput=1
\usepackage[utf8]{inputenc}
\usepackage[english]{babel}
\usepackage[T1]{fontenc}
\usepackage{amsmath}
\usepackage{hyperref}
\usepackage{amsmath}
\usepackage{amsthm}
\usepackage{bbold}
\usepackage{amssymb}
\usepackage{authblk}
\usepackage{amsfonts}
\usepackage{mathrsfs}
\usepackage{bigints}
\usepackage{braket}
\usepackage{siunitx}
\usepackage{makecell}
\usepackage{physics}
\usepackage{enumitem}
\usepackage{float}
\usepackage{blkarray}
\usepackage{braket}
\usepackage{booktabs}
\usepackage{pgfplots}
\usepackage{pgfplotstable}
\pgfplotsset{compat=1.15} 
\usepackage{multirow}
\usepackage{hyperref}
\usepackage{multicol}
\usepackage{booktabs}
\usepackage{tcolorbox}
\usepackage{lipsum}
\usepackage{xcolor}
\usepackage{wrapfig}
\usepackage{subcaption}
\usepackage{cite}
\usepackage{longtable}
\usepackage{appendix}
\usepackage{mathtools}
\usepackage{csvsimple}
\usepackage{relsize}
\usepackage{graphicx} 
\usepackage{tikz}
\usepackage{diagrammes/tikzit}
\usetikzlibrary{quantikz2}
\usetikzlibrary{matrix}

\input{commands}

\input{diagrammes/sample.tikzstyles}

\begin{document}

\title{What can we do in a symmetry-constrained perspective? The importance of the total charge's status in quantum reference frame frameworks.}

\author{Guilhem Doat}
\affiliation{LTCI, Inria, Télécom Paris, Institut Polytechnique de Paris, Palaiseau, France}
\email{guilhem.doat@telecom-paris.fr}
\author{Augustin Vanrietvelde}
\affiliation{LTCI, Inria, Télécom Paris, Institut Polytechnique de Paris, Palaiseau, France}

\maketitle

\begin{abstract}
The study of quantum reference frames has received renewed interest over the last years, leading to the parallel development of non-equivalent frameworks by different communities. We clarify the differences between these frameworks. At the mathematical level, they mainly differ in the kind of symmetry (either \textit{weak} or \textit{strong}) employed to constrain the system. We show that this mathematical difference corresponds to a fundamental physical question: whether the global charge associated to the symmetry group is accessible to symmetry-constrained observers. In this context, we formulate a definition of a perspective in terms of operational capacities, or lack thereof. Turning to consequences of adopting either approach, we discuss how adopting the weak approach induces an ambiguity in the momenta included in each perspective and bars from defining reversible QRF transformations.  We then review and analyze the existing arguments motivating each approach, and show how they bear upon the problem of charge accessibility. Finally, we introduce a simple operational scenario in which upholding two reasonable physical postulates leads to the conclusion that internal observers could measure the global charge by 1/ performing a relativized interference measurement and 2/ classically communicating.
    
\end{abstract}

\tableofcontents

\section{Introduction}

What would it mean to describe the world relative to a quantum system? This riddle presents itself as an attractive compass, pointing towards a fully relational account of quantum theory. With its inherent vagueness being a source of interest as well as of difficulty, this question has inspired significant efforts over the last decade, directed at understanding what it would mean to take it seriously. This quest has especially led to the study and use of a new type of objects: quantum reference frames (QRFs).

In particular, several frameworks have been proposed to formalize the notion of a QRF. The first attempts originate from the study of superselection rules \cite{Aharonov:1967zza, Aharonov:1984zz, Wick:1970bd}. Another approach, referred to as the quantum information (QI) approach\cite{Smith:2018gas, Spekkens:2008acy, Gour:2009hng} (see the review in \cite{Bartlett:2006tzx}), focuses on what information is ``speakable'' from within a QRF, without touching on the question of transformations from one QRF to another. The operational framework \cite{Loveridge:2016tnh, Loveridge:2017pcv, Carette:2023wpz, Glowacki:2023nnf} as well as the extra-particle framework \cite{Castro-Ruiz:2021vnq} anchor their approach in the operational meaning of QRFs, emphasizing the quantities obtainable from within a QRF. 

In parallel, the perspectival approach \cite{Giacomini:2021gei, Ballesteros:2020lgl, Giacomini:2017zju} has focused on transformations from one QRF to another. Finally, the perspective-neutral (PN) framework --  which in fact can be used to derive the former -- puts forward, using tools from constrained quantum system theory, a formalism based on a point of view containing information shared by all perspectives, hence ``perspective-neutral'' \cite{Vanrietvelde:2018pgb, delaHamette:2020dyi, delaHamette:2021oex, Hoehn:2019fsy,Hoehn:2021flk,  Krumm:2020fws, Hoehn:2023ehz}.   

These approaches have led to the discovery of many fascinating implications of the idea of QRFs, like the observer-dependent nature of entanglement or superposition \cite{Giacomini:2017zju}. Furthermore, they have enabled the application of the discoveries made within the field of QRFs to other fields. In high-energy physics, the study of edge-modes as QRFs in a quantum gravity context benefited from approaches to QRF that resemble gauge-theoretical constructions, such as the PN approach \cite{Araujo-Regado:2024dpr, Carrozza:2021gju, Carrozza:2022xut, Kabel:2023jve}. Furthermore, QRFs have also been used to propose descriptions of quantum gravity in certain new regimes, in particular superpositions of spacetimes \cite{Belenchia:2018szb, Kabel:2022cje, Kabel:2024lzr}. A final example of a successful application of QRFs to other contexts is their application to the study of indefinite causal order \cite{delaHamette:2022cka}. 

However, up until now, efforts in the field have mostly focused on deepening the understanding of each individual framework. In particular, some have been pushed to advanced mathematical contexts, in order to apply them to existing problems, like quantum gravity. It therefore remains unclear what the relationship between these various frameworks is: so far no full classification has been given, even though the non-equivalence of these frameworks is acknowledged in the literature. Most importantly, comparisons between frameworks have in majority focused on their mathematical features, and specifically on the kind of group-averaging method they use. In our view, this tackles the question from the wrong side: every mathematical assumption corresponds to a physical one, and focusing only on the former, without digging out the latter, undermines the conceptual intelligibility of the question at hand.

As a result, it remains obscure what one means when talking about a QRF. This state of affairs is of course not entirely due to the confusion around frameworks (one could even argue that the former led to the latter), but clarifying this confusion is a necessary step in pinning down precisely what a QRF is. It is also hard to understand how a problem addressed using one framework would be handled by a different one, making inter-framework dialogue tedious. Having a clear understanding of where each framework stands conceptually can only help progress on the overall meaning and significance of QRFs. Furthermore, it would allow to move from a confused overlap of approaches to a well-defined debate, in which the physical differences between the available options are made explicit.

This paper ambitions to provide such an understanding, by identifying the physical differences between frameworks, as well as adjudicating between them. We achieve this by singling out the main physical difference between frameworks as whether the total symmetry charge is accessible; and then by formulating a toy-model argument, which, from the assumption of reasonable postulates, makes us conclude that this quantity is indeed accessible. As a consequence, it seems frameworks with this feature should be preferred.

In the process of shedding light upon these questions, our contributions are the following. We first show that there are in general two different ways of imposing a symmetry on a system. We boil the difference between those down to total charge accessibility. This allows us to introduce a new terminology that hopefully clarifies the relationship between these approaches, referring to one as strong, and the other as weak. We connect this terminology with the usual distinction in terms of group-averaging techniques. We also introduce mathematical tools, namely C$^*$-algebras, as a language in which these differences become particularly explicit. Having clarified this point, we provide a critical review of the arguments in favor of each approach which exist in the literature. In all of this we restrict ourselves to the case of finite Abelian groups for simplicity, although we do not expect the conclusions we draw to be significantly different in the non-Abelian case.

Another important upshot of our work concerns the notion of perspective. We provide a definition highlighting the fact that perspectives reflect physical constraints. Mathematically, we pin them down as subalgebras. We look into the consequences that choosing each notion of symmetry has on perspectives, and most importantly on perspective changes; in particular, we show that it determines whether the latter are invertible or not. Furthermore, we showcase how within the weak approach, the content of the symmetry-constrained perspectives -- and specifically the choice of ``relative momenta'' -- is ambiguous.

Our final contribution is to propose an argument addressing the physical question of total charge accessibility on physical grounds. Postulating the operational abilities of agents in a simple toy-model using reasonable physical assumptions, we examine the physical information they can access. We find that they can collaborate to access the total charge, thus favoring the weak approach. Even irrespective of our conclusion, we see it as an important outcome that we shift the debate around QRF frameworks to the physical assumptions they (often implicitly) make, as well as the physical predictions they lead to. To our knowledge, such a physically-focused discussion had not appeared in the literature before. 

Finally, let us discuss the relationship of our contributions with previous papers that attempt to shed light on the differences between QRF frameworks. Krumm, Höhn, and Müller pioneered these discussions in \cite{Krumm:2020fws}. In the first section of this paper, they compare the QI approach and the ``structural approach'', into which they include both the perspectival and PN approaches. In doing so, they identify the main difference between these frameworks as being which group-averaging technique they use (incoherent in the QI case, and coherent in the structural case). They link these differences to the ``physicalness'' of the perspective used to distinguish symmetric states, which can also be understood as distinguishing gauge invariance from invariance under a physical symmetry. In a similar spirit, Castro-Ruiz, Galley and Loveridge provide in \cite{Castro-Ruiz:2025yvi} a neat example which they work out in a three different frameworks: PN, operational and extra-particle. This allows them to point out the key conceptual and technical differences between frameworks, as well as to showcase how each one is used. They point out the difference in charge accessibility between these frameworks. Finally another recent work aiming at comparing different QRF frameworks is by de Vuyst, Höhn, and Tsobajan \cite{DeVuyst:2025ezt}. This work proves the equivalence (at least in the ideal frames case) of three frameworks using strong symmetry (or coherent group averaging): PN, effective semi-classical \cite{Hohn:2011us, Tsobanjan:2014xqa, Bojowald:2005cw,Bojowald:2006ww, Bojowald:2009jj,Bojowald:2009jk, Bojowald:2010qw, Bojowald:2010xp, Bojowald:2009zzc}, and algebraic \cite{Bojowald:2019mas, Bojowald:2022caa, Bojowald:2022jem}.

The present work comes as complement to the works just cited. We anchor the mathematical differences highlighted in these papers into a physical one: the assumption the frameworks make on the accessibility of the total charge. We also not only state the difference in charge accessibility, as done in \cite{Castro-Ruiz:2025yvi}, but go a step further: we show that the question of charge accessibility can be elucidated through reasonable physical assumptions.

We organize the paper in the following way. In Section \ref{sec:th}, we introduce the two definitions of symmetry and connect them to the existing QRF frameworks. In Section \ref{sec:cons}, after introducing our definition of what a perspective is, we expose the consequences of choosing each approach on perspectives and, most importantly, changes of perspectives. In Section \ref{sec:lr}, we give a critical review of the existing arguments in favor of each approach to QRFs. Finally, in Section \ref{sec:oa}, we propose sensible postulates about the operational abilities of agents in a simple toy-model, and we then examine the information they can access, in order to determine which approach to favor.
We then conclude and discuss the consequences of this work for the field as well as potential future research directions.

\section{The two ways of imposing symmetry} \label{sec:th}
In this section, we introduce and clarify two different definitions of what it can mean for a system to be symmetric.
This question should not be confused with the question of how the symmetry itself is described (either by an observable, a unitary action or a group). The latter question is discussed in Appendix \ref{ann:sym}, as it complements the discussion we make here but is not essential to it. 

We discuss the physical and mathematical implications of each of these definitions, and introduce C$^*$-algebras as a formal way to make these differences explicit and tangible.

\subsection{Weak and Strong symmetry} \label{sec:w&s}
We choose to describe the state of our system with density matrices, i.e.\ positive semi-definite, linear operators of trace one on some Hilbert space $\mathcal{H}$. This is necessary as one of the approaches we want to
describe cannot be presented in a pure setting. In what follows, when we talk about an algebra of given label $i$,  we denote it as $\ca_i$, and we denote the subset of states within it using $\cs_i \subset \ca_i$. Our starting hypothesis is that the symmetry constraining our system can be described by an observable $C$, acting on the Hilbert space $\mathcal{H}$ (see Appendix \ref{ann:sym} for the connection with other ways of describing the symmetry). In accordance with Noether's seminal result, this observable is referred to as the ``total charge'' associated to the symmetry. If it is described by several charges, since we only consider Abelian symmetries, they all commute and can then be summed into one single one, which we will then regard as the charge operator $C$. In what follows, we will see that the relevant data is in fact given by the eigenspaces of this operator $C$ so we could also choose any given (non-degenerate) function of it.

For simplicity, we will only consider discrete abelian symmetries, and in this context working with states is dual to working with observables \cite{Baez:2011hu}. In what follows, our goal is to identify what is the relevant subset of symmetric states $\cs_{\text{sym}} \subset \mathcal{S}(\mathcal{H})$. Physically, this means that we are looking for the subset of states, or equivalently observables, that respect the symmetry. How precisely this subset is identified depends on important physical assumptions, which we highlight below. A natural example is the following.

\begin{example} \label{ex:1}
Consider two systems $A$ and $B$ living on a one-dimensional circular lattice with $3$ sites. The symmetry transformation is described by the $3$-elements cyclic group $G=\mathbb{Z}_n = ( \{0,1,2\} , \oplus \equiv +\mod 3)$, as each system's position can take any of these values. We can describe the state of the global system using each individual system's position, and as a consequence the total Hilbert space $\ch_A \otimes \ch_B$  is spanned by the family $\{ \ket{i,j}\}_{i,j=0,1,2}$, which we refer to as the position basis. The observable associated to the symmetry here is the \textit{pseudo-momentum} $C$, generating translations in the position basis as
\be
e^{iC} \ket{i,j} = \ket{i\oplus 1, j \oplus 1}.
\ee
\end{example}

This operator $C \in \textrm{Lin}(\mathcal{H})$ can then be diagonalized and $\mathcal{H}$ sectorized into its eigenspaces, which we shall call \textit{charge sectors} of the charge observable $C$. If $C$ has $L$ different eigenvalues $\lambda_l$, we write
\be
\mathcal{H} = \bigoplus_{l=1}^{L} \mathcal{H}_{l}.
\ee
 This sectorization also outlines a suitable representation of the charge observable $C$ as a matrix, which in its own  eigenbasis is of course diagonal:
\be
C= \text{diag}(\lambda_1,...,\lambda_1,\lambda_2,..., \lambda_2,......,\lambda_l,..., \lambda_l).
\ee
By convention, we take $\lambda_1 = 0$ whenever $0 \in \text{Spec}(C)$.

We now understand how the Hilbert space associated with our problem sectorizes under the action of the charge observable; we now want to identify the subset of symmetric operators $\ca_{\text{sym}}$ from there. The first central point of the present paper is that there are two unequivalent approaches we can follow:
\begin{itemize}
    \item \textbf{Weak approach:} We define $\ca_{\text{sym}}$ as the subset of operators that commute with the charge observable $C$. Using the diagonal form of the charge operator, we can infer that operators commute with $C$ if and only if they are block diagonal operators with respect to the sectorisation into charge eigenspaces. We can then write 
    \be
    \ca_{\text{sym}} = \bigoplus_l \text{Lin} (\mathcal{H}_{l}) \equiv \ca_{W} = \{A \in \text{Lin}(\ch) | [C, A]= 0 \},
    \ee
    the space of matrices of the form
    \[
    \begin{tikzpicture}
    \matrix (m1)    [matrix of nodes,
                     left delimiter={(}, right delimiter={)},
                     row sep=-0.5pt,column sep=-0.5pt,
                     every node/.style={inner sep=5pt},
                     F/.style = {draw=red, line width=0.5pt, fill=red!20}
                     ]
    {
    |[F]| \color{red!20} $\lambda_0$  &   \ldots          &   $\mathbf{0}$             \\
     \vdots         & |[F]| $\ddots$  &   \vdots           \\
     $\mathbf{0}$       &   \ldots           & |[F]| \color{red!20} $\lambda_0$ \\
     };
        \end{tikzpicture},
    \]
    where the blocks highlighted in red are those allowed to contain non-zero values.
    Our convention is that the top-left block corresponds to the zero-charge subspace.
    \item \textbf{Strong approach:} We define $\ca_{\text{sym}}$ as 
    \be
    \ca_{\text{sym}}=\text{Lin}(\mathcal{H}_{0}) \equiv \ca_{\text{S}}= \{ A \in \text{Lin}(\ch) | \text{supp}(A) \subseteq \ch_0 \} \, ,
    \ee
    which in matrix form translates to the structure
    \[
    \begin{tikzpicture}
    \matrix (m1)    [matrix of nodes,
                     left delimiter={(}, right delimiter={)},
                     row sep=-0.5pt,column sep=-0.5pt,
                     every node/.style={inner sep=5pt},
                     F/.style = {draw=red, line width=0.5pt, fill=red!20}
                     ]
    {
    |[F]| \color{red!20} $\lambda_0$ &   \ldots          &   $\mathbf{0}$             \\
     \vdots         & $\mathbf{0}$  &   \vdots           \\
     $\mathbf{0}$       &   \ldots           & $\mathbf{0}$   \\
     };
        \end{tikzpicture}.
    \]
\end{itemize}
 Thus, choosing the weak approach means restricting to block-diagonal density matrices in the charge sectorization, while the strong option means restricting to a subset of those, with  non-null coefficients only in the top-left block. 
 
 Turning to the physical meaning of this, imposing invariance under weak symmetry means forbidding superpositions of charge eigenstates, i.e.\ banning coherence between charge sectors. Imposing strong symmetry, on the other hand, means enforcing the charge to be null. This is also where these approaches take their respective names: imposing the weak approach is a weaker restriction than the strong approach since $\ca_{\text{S}} \subseteq \ca_{\text{W}}$.

In the literature, in general, the main point of discussion to differentiate approaches is which projector they use; let us connect with it. To this end, let us consider a physical system constrained by a symmetry, described by a finite abelian group $G$ of order $n$. The action on this group on the Hilbert space of the system $\ch$ is a unitary representation $U: G \mapsto \textrm{Lin}(\mathcal{H})$.
 The projector onto the block-diagonal states in the charge eigenbasis, which we call the \textit{weak twirl}, is sometimes referred to as incoherent group averaging, and can be expressed as 
 \be \label{eq:tw}
 \begin{split}
     \mathcal{T}_W  &: \textrm{Lin} \left (\mathcal{H} = \bigoplus_{l=1}^L \mathcal{H}_l \right )  \mapsto \bigoplus_{l=1}^L \textrm{Lin} \left (\mathcal{H}_l  \right )\\
     \rho &\rightarrow \frac{1}{n} \sum_{g \in G} U(g) \rho U^{\dagger}(g).
 \end{split}
 \ee 
 On the other hand, the projector onto the zero-charge sector, which we shall call the \textit{strong twirl}, also referred to as coherent group averaging, can be written as 
 \be
 \begin{split} \label{eq:ts}
     \mathcal{T}_S  &: \textrm{Lin} \left (\mathcal{H} = \bigoplus_{l=1}^L \mathcal{H}_l \right )  \mapsto \textrm{Lin} \left ( \mathcal{H}_0 \right ) \\
     \rho &\rightarrow \frac{1}{n^2} \sum_{g,g'\in G} U(g) \rho U^{\dagger}(g').
 \end{split}
 \ee
We would like to emphasize, however, that this form of the projector tends to obfuscate the fact that their action, when $\rho$ is seen in the appropriate matrix form, is simply to ``kill'' $\rho$'s blocks outside the chosen $\ca_{\text{sym}}$. We think this is a simpler way of thinking about the two different approaches. What truly sets them apart is the physical content of the symmetric algebra, reflected in the choice of $\ca_{\text{sym}}$.

\begin{example} \label{ex:2}
Consider again two systems A and B living on a symmetric three-sites lattice. We now want to find the exact form of the pseudo-momentum $C$, as well as sketch the structure of the symmetric algebra $\ca_{sym}$ in the weak and strong approaches. 

The system is described by $\rho_{AB} \in \cs (\ch =\mathcal{H}_A \otimes \mathcal{H}_B )$, where the tensor factors are three-dimensional spaces. The $\mathbb{Z}_3$ symmetry on this total system is realized by the global shift operator $U^{\otimes 2}$.
This global translation operator $U^{\otimes 2} \in \textrm{Lin} (\mathcal{H})$ can be diagonalized to find the different charge sectors.  Since applying it three times gives back the same state of the whole system \textit{i.e.} $(U^{\otimes 2})^3=  \mathbb{1}$, the eigenvalues of $U^{\otimes 2}$ are given by roots of unity. The associated eigenstates are (see \cite{Krumm:2020fws}, Example 8)
\be \label{eq:es}
\ket{c;r} = \frac{1}{\sqrt{3}} \sum_{n=0}^2 e^{-\frac{2i \pi}{3} n c} \ket{n,~ n\oplus r}
\ee
where $c = 0,1,2$ is the charge and $r = 0,1,2$ the relative distance between the two systems. Here, the $r$ index indicates a degeneracy of each charge value $c$. States within a charge sector labeled by $c$ that only differ by their index $r$ transform the same under the global translation, and so do their superpositions. In other words, the eigenstates of the shift operator, which generates the symmetry, are all the possible superpositions of states that share the same relative positions. Each of these states pick up a global phase under $U^{\otimes2}$:
\be
U^{\otimes2} \ket{c;r} = e^{\frac{2 i \pi}{3}c} \ket{c;r}.
\ee
Thus, the spectrum of $U^{\otimes 2}$ is $ \sigma = \{ 1, e^{\frac{2 i \pi}{3}},e^{\frac{4 i \pi}{3}} \}$. We also see that the eigenspace associated to each eigenvalue is of dimension 3. In consequence, we can write the diagonal form of $U^{\otimes 2}$, i.e.\ in its eigenbasis we have
 \be
 U^{\otimes 2}= \textrm{diag} (1,1,1,e^{\frac{2 i \pi}{3}},e^{\frac{2 i \pi}{3}},e^{\frac{2 i \pi}{3}},e^{\frac{4 i \pi}{3}}, e^{\frac{4 i \pi}{3}}, e^{\frac{4 i \pi}{3}}).
\ee 
As we discussed in Appendix \ref{ann:sym} and Example \ref{ex:1}, the charge observable $C$ associated to $U^{\otimes 2}$ through the exponential map is indeed given by 
\be
C= \frac{2\pi}{3}\cdot\textrm{diag}(0,0,0,1,1,1,2,2,2).
\ee
Calling the eigenvalue index $c$ the charge, we can see that the top-left block is  the zero-charge sector.
This observable can be thought of as a ``momentum'', since it plays a similar structural role to momentum in the case of translation symmetry: it generates the elementary symmetry transformation, analogously to the way the momentum generates  infinitesimal translations.
As a consequence of this eigenstructure, the Hilbert space partition takes the form 
\be
\mathcal{H} = \mathcal{H}_0 \oplus \mathcal{H}_1 \oplus \mathcal{H}_2.
\ee
$\mathcal{H}_{0}$  is the zero-charge subspace, i.e.\ the invariant subspace under the action of the group. $\ch_{1}$ and $\mathcal{H}_{2}$ are the space of states which transform respectively with a $e^{\frac{2 i \pi}{3}}$ or $e^{\frac{4 i \pi}{3}}$ phase under the action of the group:
\be
\begin{split}
    \mathcal{H}_0 &= \text{Span} \{ \ket{0; r}, r \in \Z_3 \} \\
    \mathcal{H}_1 &= \text{Span} \{  \ket{1; r}, r \in \Z_3 \} \\
    \mathcal{H}_2 &= \text{Span} \{  \ket{2; r}, r \in \Z_3 \}.
\end{split}
\ee

As we explained in this section, imposing strong symmetry on this system means considering $\cs_{\text{sym}}=\mathcal{S}(\mathcal{H}_{0})$, i.e.\ states of the form
\be
\rho_S = \sum_{r,r'\in \Z_3} \alpha_{r,r'} \ketbra{0;r}{0;r'}
 = \begin{bmatrix}
    \alpha_{00} & \alpha_{01} & \alpha_{02} & 0 & 0 & 0 & 0 & 0 & 0 \\
    \alpha_{10} & \alpha_{11} & \alpha_{12} & 0 & 0 & 0 & 0 & 0 & 0 \\
    \alpha_{20} & \alpha_{21} & \alpha_{22} & 0 & 0 & 0 & 0 & 0 & 0 \\
    0 & 0 & 0 & 0 & 0 & 0 & 0 & 0 & 0 \\
    0 & 0 & 0 & 0 & 0 & 0 & 0 & 0 & 0 \\
    0 & 0 & 0 & 0 & 0 & 0 & 0 & 0 & 0 \\
    0 & 0 & 0 & 0 & 0 & 0 & 0 & 0 & 0 \\
    0 & 0 & 0 & 0 & 0 & 0 & 0 & 0 & 0 \\
    0 & 0 & 0 & 0 & 0 & 0 & 0 & 0 & 0
\end{bmatrix}, ~~ \alpha_{r,r'} \in \C.  
\end{equation}
and, on the other hand, imposing weak symmetry means setting $\cs_{\text{sym}} \cong \mathcal{S}(\mathcal{H}_{0})\oplus \mathcal{S}(\mathcal{H}_{1}) \oplus \mathcal{S}(\mathcal{H}_{2})$, i.e.\ states of the form
\be
\rho_W = \sum_{r,r',c} \alpha_{r,r';c} \ketbra{c;r}{c;r'}
= \begin{bmatrix}
    \alpha_{00;0} & \alpha_{01;0} & \alpha_{02;0} & 0 & 0 & 0 & 0 & 0 & 0 \\
    \alpha_{10;0} & \alpha_{11;0} & \alpha_{12;0} & 0 & 0 & 0 & 0 & 0 & 0 \\
    \alpha_{20;0} & \alpha_{21;0} & \alpha_{22;0} & 0 & 0 & 0 & 0 & 0 & 0 \\
    0 & 0 & 0 & \alpha_{00;1} & \alpha_{01;1} & \alpha_{02;1} & 0 & 0 & 0 \\
    0 & 0 & 0 & \alpha_{10;1} & \alpha_{11;1} & \alpha_{12;1} & 0 & 0 & 0 \\
    0 & 0 & 0 & \alpha_{20;1} & \alpha_{21;1} & \alpha_{22;1} & 0 & 0 & 0 \\
    0 & 0 & 0 & 0 & 0 & 0 & \alpha_{00;2} & \alpha_{01;2} & \alpha_{02;2} \\
    0 & 0 & 0 & 0 & 0 & 0 & \alpha_{10;2} & \alpha_{11;2} & \alpha_{12;2} \\
    0 & 0 & 0 & 0 & 0 & 0 & \alpha_{20;2} & \alpha_{21;2} & \alpha_{22;2}
    \end{bmatrix}.
\end{equation}
These states only allow coherence within each charge block, and forbid coherence between them.
\end{example}

\subsection{Mathematical features of the two approaches: factoriality and non-factoriality } \label{sec:c*}

We now present the mathematical properties that characterize the two definitions of symmetry previously discussed, as they allow for a deeper comprehension of their structural differences. Since the different choices of $\ca_{\text{sym}}$ are all C*-algebras, introducing a general theory of the latter will prove handy. Let us start with a basic definition. 
\begin{definition}
    A finite dimensional *-algebra $\ca$ is a finite-dimensional algebra over $\C$ equipped with an involution $\dagger$ which respects the algebra's structure, i.e.\ for any three elements $A_1, A_2$ and any scalar $\lambda$ we have $(A_1+A_2)^{\dag} = A_1^{\dag}+A_2^{\dag}$, $(A_1A_2)^{\dag} = A_2^{\dag}A_1^{\dag}$ and $(\lambda A_1)^{\dag} = \lambda^*A_1^{\dag}$. A sub*-algebra is a subalgebra of $\ca$ that is closed under the dagger.
    
    Furthermore, we can define a finite-dimensional C*-algebra as a *-algebra that is isomorphic to some sub*-algebra of $\textrm{Lin}(\ch)$, where $\ch$ is a Hilbert space, and the involution is the adjoint associated to its inner product. Any sub*-algebra of a C*-algebra is then a C*-algebra, and can thus be called a sub-C*-algebra.
\end{definition}

To see where we are going with these rather blunt definitions, let us state an important theorem to help build a good intuition of what C*-algebra look like (in finite dimension). This theorem, due to Wedderburn and Artin, states that, in general, a C* algebra is not always isomorphic to an algebra of operators $\textrm{Lin}(\ch)$ but to a direct sum of such algebras.
\begin{theorem}
    (Wedderburn-Artin) Given a finite dimensional C* algebra $\ca$, there exists a list of finite-dimensional Hilbert spaces $(\ch_l)_{k\in L}$, indexed from 1 to L, such that
    \be
    \ca \cong \bigoplus_{l=1}^L \textrm{Lin}(\ch_l).
    \ee
\end{theorem}
This result provides an intuitive way to understand C*-algebras. In finite dimension, they are nothing more than an algebra of block-diagonal operators over a single Hilbert  space $\ch$ with a preferred sectorization $\ch = \bigoplus_l \ch_l$, and we can write
\be
\ca \cong \left \{ f \in \textrm{Lin}(\ch) | f(\ch_l) \subseteq \ch_l,~  \forall l \right \}.
\ee
A C*-algebra that is isomorphic to a single $\textrm{Lin}(\ch)$ is called a \textit{factor}, whereas one that is only isomorphic to a direct sum of such algebras is a \textit{non-factor}. 
In the context of quantum theory, one can understand non-factorness as ``partial classicality''.  Being block diagonal essentially means that there can be no coherence between the eigenspaces (the sectors) of a certain observable. All we can then observe are classical mixtures of the possible outcomes a measurement of this given observable can yield.

This theory allows us to highlight a key difference between our two candidates for $\ca_{\text{sym}}$. The algebra of weakly symmetric states, $\ca_{\text{W}}$, is as we have seen an algebra of block diagonal matrices in the charge eigenbasis, and thus a non-factor one, whereas $\ca_{\text{S}}$, the algebra of strongly symmetric states, is a factor. A first basic conclusion we can draw from this is that the two different approaches yield two algebras of states $\ca_{\text{W}}$ and $\ca_{\text{S}}$ that are not isomorphic due to their non-factor/factor nature. This sharp difference at the level of their mathematical structure is a strong indicator that the two approaches are non-equivalent at the physical level. The physical intuition behind this mathematical difference is that $\ca_{\text{W}}$ contains more observables than $\ca_{\text{S}}$, and in particular one more classical (meaning that it commutes with all the elements in $\ca_{\text{W}}$) observable: the charge. 

For clarity, we summarize the differences presented in this section in Table \ref{tab}.

\begin{table}[h]
    \centering
    \renewcommand{\arraystretch}{1.3} 
    \begin{tabular}{|m{5cm}|m{5cm}|m{5cm}|}
        \hline
        \makecell[c]{\rule{0pt}{1.5em}} & \makecell[c]{\textbf{Strong approach}} & \makecell[c]{\textbf{Weak approach}} \\ \hline
        \makecell[c]{Symmetry \\ criterion} 
        & \makecell[c]{Annihilation: \\ $C\ket{\psi} = 0$} 
        & \makecell[c]{Commutation: \\ $[C, \rho] = 0$} \\ \hline
        
        \makecell[c]{Structure of the \\ symmetric algebra \\ $\mathcal{A}_{\text{sym}}$}
        & \makecell[c]{\raisebox{-.5\height}{\includegraphics[width=0.57\linewidth]{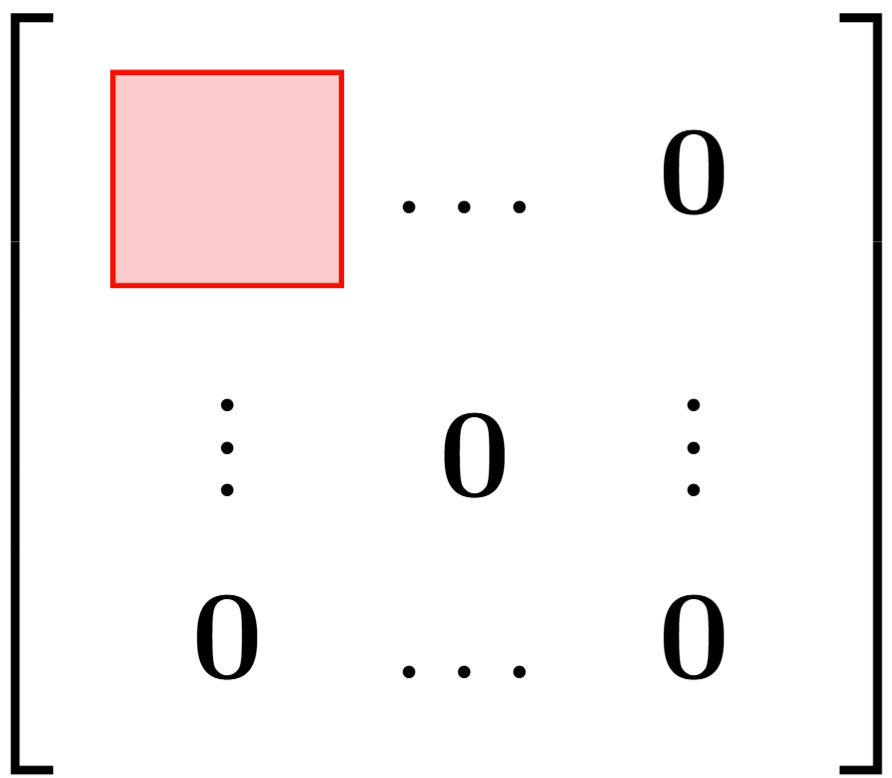}}}
        & \makecell[c]{\raisebox{-.5\height}{\includegraphics[width=0.6\linewidth]{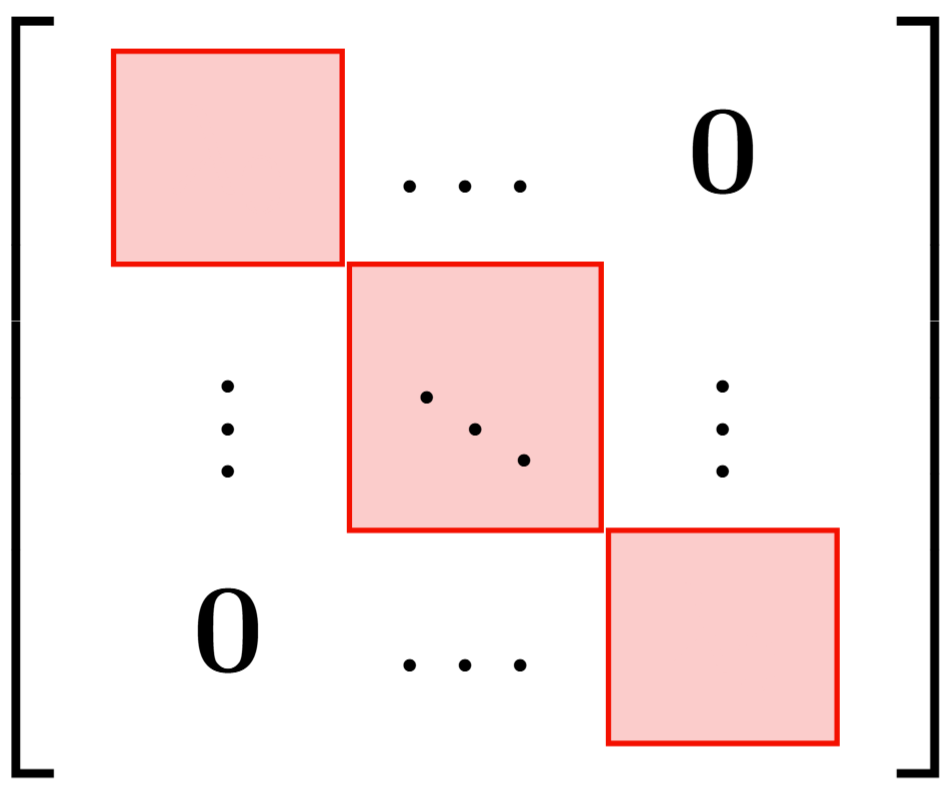}}} \\ \hline
        
        \makecell[c]{Reversibility of \\ QRF transformations} 
        & \makecell[c]{Yes} 
        & \makecell[c]{No} \\ \hline
        
        \makecell[c]{Accessibility of \\ the total charge} 
        & \makecell[c]{No} 
        & \makecell[c]{Yes} \\ \hline
        
        \makecell[c]{Projector \\ (Twirl)} \rule{0pt}{1.5em} 
        & \makecell[c]{$\frac{1}{n^2} \sum_{g,h \in G} U(g) \rho U^{\dagger}(h)$} 
        & \makecell[c]{$\frac{1}{n} \sum_{g \in G} U(g) \rho U^{\dagger}(g)$} \\ \hline
        
        \makecell[c]{Nature of \\ $\mathcal{A}_{\text{sym}}$} 
        & \makecell[c]{Factor} 
        & \makecell[c]{Non-factor} \\ \hline
    \end{tabular}
    \caption{\textit{Summary of the differences between the approaches.}}
    \label{tab}
\end{table}

\section{Perspectives in the weak and strong approaches} \label{sec:cons}
The goal of this section is, building on the previous one, to introduce the notion of perspective. We aim to be precise and consistent with the terminology used in order to avoid confusion between the words we employ, and words that have previously been proposed in the literature. Inevitably, the definitions we give involves a certain number of assumptions, which we clarify and justify. Having done this, we move on to describing the consequences that choosing either the strong or the weak approach has on the physical quantities accessible from each perspective. This will showcase how the different mathematical tools chosen to build the theory affect the physical content ascribed to perspectives.

\subsection{What is a perspective?} \label{sec:agents}
First, let us clarify  what we mean when we talk about perspectives. This concept is central to the field of QRFs as whole, and of course to our development; in fact, it is the very object of study we want to understand. Nevertheless, it has been given many definitions throughout the literature, and it is unclear how they all relate. The word perspective is also often used as a synonym of QRF, but here we make a clear distinction between those concepts. We begin with our definition.

\begin{definition} \label{def:persp}
A perspective is 1/ a restriction on operational abilities, and 2/ a specific labeling of available operations, reflecting the constraint of only using specific physical means to interact with the world.
\end{definition}

For instance, a QRF is a specific type of perspective in which ``the constraint of only using specific physical means'' means restricting to only act via the interactions of a single subsystem with all others. The more general concept of perspective encompasses a wider range of situations, in which the restriction does not necessarily come from the restriction of interacting through a single subsystem.

Mathematically, we will assume that a perspective corresponds to a subalgebra of $\text{Lin}(\ch)$ where $\ch$ is the total Hilbert space of the theory.  This connects with the approach developed in \cite{Castro-Ruiz:2021vnq}, where the authors derive results on accessible quantities in different QRFs using the fact that the algebras associated to them overlap within $\text{Lin}(\ch)$. While the authors had pinned down the relevance of subalgebras to describe perspectives, we add to their approach the precise idea that perspectives arise from physical restrictions. 

This assumption is also based on the ideas developed in the theory of  partitions of quantum systems, proposed in \cite{Vanrietvelde:2025tjn}. In this work, the authors discuss the consequences of associating algebras to perspectives, as well as where this idea stems from. They also include non-factor algebras to their analysis, which is useful in our case, as presented in Section \ref{sec:c*}.

Operationally, what we mean by associating an algebra $\ca$ to a perspective can be interpreted in three equivalent ways: 1/ the Hermitian operators in $\ca$ describe the observables that an agent adopting the perspective can measure; 2/ the unitaries in $\ca$ describe the evolutions she can implement; or 3/ the operators in $\ca$ are the possible Kraus operators of the channels she can apply. See the discussion in Section 2.1 of \cite{Vanrietvelde:2025tjn} as well as \cite{Ormrod:2025dax}.

A crucial question to ask about Definition \ref{def:persp} is of course where the said restriction might originate from. The two possibilities are 1/ an \textit{in-principle} restriction -- i.e.\ a restriction that an agent could not overcome even if she wanted to -- or 2/ a merely \textit{methodological} restriction -- i.e.\ a restriction that an agent, out of interest for what would happen if she were to be subject to it, self-imposes, in her theoretical description and/or in her experimental practice. A natural stance, in the face of this dichotomy, would be to only attribute foundational relevance to in-principle restrictions, and to treat methodological restrictions as, at best, coarse-grainings with little such relevance.\footnote{Indeed, a similar stance has commonly been adopted \cite{Vilasini2022, Ormrod2023, Vilasini2025} regarding the -- very related, but this is another story -- question of whether it can be said that indefinite causal order has been realized in the lab, or merely simulated. For previous pushback against this stance in that context, see \cite{oreshkov2019time, delaHamette:2022cka}.} However, we want to push against such a stance here.

Our main pushback argument is that no perspective-related restriction can be fully in-principle. Indeed, either this restriction holds at any level of description of the universe, for any agent -- in which case it is not a restriction but simply the global set of possible operations (since the latter includes nothing outside of it) and the perspective at hand is no different from the most powerful conceivable perspective, except maybe in its labelling of operations; or there exists a level of description at which this restriction is lifted -- in which case we expect that it is always conceivable for any agent to communicate with another agent acting at this level, so that for any agent, restricting to the perspective is a merely methodological choice.\footnote{There is an interesting parallel to be drawn with the status of gauge symmetries in physical theories -- a subject of constant philosophical debate. It is usually held that a gauge symmetry is `fully in-principle'; however, this position crucially relies on the \textit{cosmological assumption}, i.e.\ the idea that the theory at hand models the entire universe \cite{Wallace2022}. If instead we interpret the theory as modeling a subsystem of a larger reality, then operational restrictions due to gauge symmetry (of the subsystem) can typically be partially overcome by acting on a larger system, and the only `fully in-principle' restriction is the one corresponding to gauge symmetry of the entire universe. This fact has indeed been proposed as an explanation for why gauge theories are heuristically so powerful, when one may naively believe that they just bring in superfluous, meaningless degrees of freedom \cite{Rovelli2014, Francois2024}. (This is only a parallel, however: the perspectival restrictions of Definition \ref{def:persp} are due not to a gauge symmetry, but to the restriction to specific physical means of interaction -- even though the \textit{effects} of this restriction will typically be affected by the existence of a gauge symmetry.)} Thus, to the extent that we want to confer foundational relevance to \textit{any} notion of perspective, our view is that we have to accept the fact that the restrictions coming with it can always be seen as methodological.\footnote{The other choice is to simply deny foundational relevance to perspectives at all; in which view QRFs, in particular, hold little interest whatsoever.} A crucial point for us, securing the foundational relevance of the use of perspectives (and in particular of QRFs), is thus the following: restricting to a perspective as a methodological tool is  legitimate, as long as there is no reason to forbid the conceiving of an agent from whose point of view the description is natural. Said a bit more informally,
\begin{center}
\textit{role-playing is OK}.
\end{center}

By this we mean that if it is conceivable, without affecting the physics of interest, that very complex ancillary systems -- even perhaps amounting to a full lab -- may exist for which this perspective is natural,  then it is physically legitimate, and sometimes foundationally relevant, to talk about this perspective in general, and in particular to enact it -- in order, for instance, to experimentally probe theoretical predictions about it -- by artificially restricting our operations to those accessible to agents that would exist within these very complex ancillary systems, i.e., by role-playing these agents.
This is why we are able to identify agents with the physical subsystem to which perspective they restrict.

Let us take an example in the case of a simple instance of a QRF, the one attached to a quantum point-like object, potentially in a superposition of positions. It is possible, without affecting the physics of interest, to endow this system with additional internal degrees of freedom, and nothing forbids them from being so complex that they amount to a full laboratory. Our main claim, in that context, is that if this laboratory could exist in principle, then it is legitimate to talk about the ``perspective'' it might have on the world.

We expect this claim, which is crucial to the relevance and interpretation of the notion of a physical perspective, to be challenged; let us preemptively defend it by taking an example of similar reasoning from an uncontroversial part of physics. In special relativity, to illustrate the time dilation effect predicted by the theory, physicists colloquially refer to a pair of agents measuring different time intervals between two events, depending on their relative speed. Of course, this construction is, as of now, purely methodological: so far, no agents (construed, for instance, as human beings) have experienced time dilation ``themselves''. The point is that the existence of an agent corresponding to a physical system is a potentiality, as it is allowed by the laws of physics; so that it is legitimate, and fruitful, to talk about such an agent's perspective.

Let us make this more precise by discussing a typical empirical confirmation of time dilation; the sea-level detection of atmospheric muons.  (As a reminder, this phenomenon is understood via reasoning in the reference frame where the muon is at rest: in this frame, the time it takes it to reach the ground is short enough for the muon not to disintegrate.) It goes without saying that no muon-riding agent is needed to claim that this is a proof of time dilation. When we say we describe physics ``relative to the muon'', we mean that we include in  the description only the degrees of freedom of the muon in its own reference frame. We then can, and often do, perform the illustrative leap to say that ``to the muon, time flows differently'' because it would be perfectly compatible with the laws of physics to attach a laboratory to the muon to really see the world from its perspective and because, as shown by the non-disintegration of the muon at sea level, such a laboratory would experience time dilation relative to us. 

To summarize and conclude, it is not necessarily illegitimate to enforce methodologically the restriction to a limited subset of possible operations. Beyond its legitimacy, this restriction also holds explanatory power; for instance, even though time-dilation can be stated at the raw, abstract level of Lorentz transformations, it is much more intelligible when presented more concretely in terms of clocks accessible to a moving observer.

\subsection{Global and collaborative perspectives} \label{sec:persp}

Following this delineation of our notion of perspective, we present, in the context of symmetry-constrained systems as introduced before, a short bestiary of relevant perspectives -- mathematically defined through the subalgebras they correspond to -- and terminology we will use. 

We denote by $\ca_{\text{kin}}$ the kinematical algebra. We associate this algebra to the external perspective. This algebra serves as a theoretical starting point, in which we can distinguish physical quantities which we will later want to call symmetric. As such, this construction can be associated to an-all powerful observer, not subject to any constraint. 

$\ca_{\text{sym}}$ is then the symmetric algebra, i.e.\ $\ca_{\text{sym}} \subseteq \ca_{\text{kin}}$ is the subset of operators satisfying the chosen symmetry condition, and we refer to those as the symmetric operators. It is is associated to a projector $\ct: \ca_{\text{kin}} \mapsto \ca_{\text{sym}}$, which is the twirl that has been the focus of much of the literature. The elements of $\ca_{\text{sym}}$ are thus invariant under the twirl's action. As discussed in the  previous section, in the existing QRF literature, $\ca_{\text{sym}}$ has been taken to be either $\ca_{\text{W}}$ or $\ca_{\text{S}}$, which entails different physical consequences.

Given a tripartite system $ABC$,  $\ca_{BC|A}$ is the perspectival algebra associated to A. It contains the operators describing the operations  that an agent adopting the point of view of A has access to to act on the bipartite system $BC$. What exactly perspectival algebras should contain has been up for debate throughout the literature, and we will present the different stances regarding this matter later.

We call ``collaborative perspective'' the algebraic span of the union of the perspectival algebras $\ca_{\text{col}}$ = $\ca_{AB|C} \vee \ca_{AC|B} \vee \ca_{BC|A}$. Physically, it contains all the information $A$, $B$ and $C$ can access if they team up. They could for example exchange some ancilla,  on which they would be able to inscribe the result of their measurements. This algebra is, to the best of our knowledge, never mentioned in the QRF literature, and we want to stress the following crucial fact about it: it can be different from $\ca_{\text{sym}}$. This difference is central to our argument, as we will  discuss below. The content of the perspectival algebras will naturally determine the content of the collaborative algebra; therefore, the debate we were just mentioning in the previous paragraph also has implications for $\ca_{\text{col}}$.

To make things explicit, let us consider the familiar example of the translation group. Here $\ca_{\text{kin}}$ would be the algebra of all possible operators on the Hilbert space associated with the system $L^2(\RR)$, including those that are not translation-invariant. To get the symmetric algebra $\ca_{\text{sym}}$, we select only states that are translation-invariant. As one can anticipate from our discussion earlier, imposing translation invariance can here mean two things; either we select $\ca_{\text{sym}} = \ca_{\text{W}}$, and forbid coherence in momentum, or we choose $\ca_{\text{sym}} = \ca_{\text{S}}$, and keep only operators with zero momentum.

Now, in this example, what should $A$'s perspectival algebra contain? In our translation group case, we are interested in describing position QRFs, so it should at least contain relative positions. Therefore we have $x_B-x_A, x_C-x_A \in \ca_{BC|A}$. We could stop there, and have a fully classical (non-factor) algebra. If we want to claim  to describe ``quantum'' reference frames, however, we also want $\ca_{BC|A}$ to contain observables that correspond to information on the potential coherence between eigenstates of positions of $B$ and $C$ relative to $A$; in other words,  momentum observables. In general, in our view, the point of a \textit{quantum} reference frame, as opposed to a classical one, is that we do not want any operator that would be classical from the point of view of the perspectival algebra, i.e.\ that would commute with all its elements. Thus $\ca_{BC|A}$ must have a trivial center, or in other words be a factor. In particular, it does not contain the total charge, which of course would commute with the entire algebra. The natural question is then: which momenta should the perspectival algebras contain? This is a central point of this paper, and as we shall see, there is not necessarily a straightforward answer to that question.

\subsection{Perspectives in the weak approach} \label{eq:weakpersp}
Let us start by describing perspectives in the weak approach.  Our aim is to present the physical implications of this choice, and to expose all the assumptions one makes doing it. Most importantly, we explain why opting for the weak approach results in non-invertible perspective transformations. We also detail how the weak approach results in an ambiguity in the momenta operators contained in the symmetric algebra $\ca_{\text{sym}}$. Finally, we present how the relativization map usually used in the weak approach makes a choice of relative momenta which does not appear to be justified.

As previously stated, choosing $\ca_{\text{sym}} = \ca_{\text{W}}$ implies that there can be no isomorphism linking $\ca_{\text{sym}}$ to any perspectival algebra, since $\ca_{\text{sym}}$ is a non-factor algebra, and all the perspectives are factors. The physical counterpart to this mathematical fact, is that the charge information, i.e.\ which charge sector an operator belongs to, is in $\ca_{\text{W}}$, but not in $\ca_{BC|A}$. More precisely, the operators in $\ca_{BC|A}$ are mixed in the total charge: they are a sum of the same operator on each of the charge sectors.
    
A first important consequence of going with the weak approach is momenta ambiguity. The charge information is, as we just stated, in $\ca_{\text{sym}}$. Furthermore, it is in the center of this algebra, meaning it commutes with all of its elements. As a consequence, any multiple of the charge operator can be added to the elements of $\ca_{\text{sym}}$, without changing the commutation relations. This means that there is some degree of ambiguity in what the perspectival algebras, which are included in $\ca_{\text{sym}}$, can contain since to any operator in these algebras we can add elements from the non-trivial center of $\ca_{\text{sym}}$, without changing any commutation relation. To see what consequences this can have, let us look at an example.  In the case of the translation group, the charge is the total momentum. Since $\ca_{\text{sym}}$ is the algebra of observables commuting with $P_{\text{tot}}$, we can always redefine our momenta like: $P_i \mapsto P_i + \alpha P_{\text{tot}}$ because this does not modify the commutation relations, leading to different possible definitions of each perspectival algebra. Let us now consider a bipartite scenario in which we would like to know what momentum each perspectival algebra contains. Two  allowed choices would be $\{P_1, P_2\}$ and $\{P_{r1}=1/2(P_1-P_2), P_{r2} = - P_{r1}\}$, as they are linked through $P_{i} \mapsto P_{ri} = P_i - 1/2 P_{\text{tot}},~ P_{\text{tot}} =P_1+P_2$. In fact, different papers in the literature have adopted each of these sets: \cite{Angelo_2011} chooses to use the algebra containing  $\{P_{r1}=1/2(P_1-P_2), P_{r2} = - P_{r1}\}$, whereas others, such as \cite{Castro-Ruiz:2021vnq}, choose $\{P_1, P_2\}$ instead.
     
There is a second ambiguity in the perspective transformations. Imposing the weak approach also implies that there is no natural way to define an isomorphism from one perspectival algebra to another since one cannot simply change perspective through $\ca_{\text{sym}}$. The easiest way to see this is to examine the content of both the perspectival algebras and the symmetric algebra. As mentioned in Section \ref{sec:c*}, the weakly symmetric algebra is not a factor since it contains the total charge, while the perspectival algebras are factors. From there, we are able to conclude that these cannot be isomorphic to one another. Physically, this means that choosing the weak approach to impose symmetry will yield perspectives that contain different information. By this we do not simply mean that they contain different descriptions of the same information, but that not all internal observer can infer the same predictions about the observed system. Given the perspectival algebra associated to a subsystem, an agent adopting this perspective cannot fully reconstruct what she would see if she adopted the perspective of another subsystem.

From the two ambiguities described above stems a third one; what does $\ca_{\text{col}}$ contain? As hinted in the previous section, the collaborative algebra $\ca_{\text{col}}$ is different from $\ca_{\text{sym}}$ in general. Indeed, since the content of the perspectival algebras is ambiguous, so is the algebraic span of their union. It can then happen that this span fully recovers the symmetric algebra $\ca_{\text{sym}}$, but also that some information is still missing. This indeed happens in our translation group example: the two previously presented allowed sets of perspectival momenta entail different collaborative algebras. The total momentum $P_{\text{tot}}$ is in the algebraic span of $\{P_1, P_2\}$, but not in that of $\{P_{r1}=1/2(P_1-P_2), P_{r2} = - P_{r1}\}$. Therefore, the agents adopting each perspective would be able to collaborate to get the value of the total charge $P_1+P_2$ in one case, thus resolving the full symmetric algebra, and not in the other.  This ambiguity has recently been mentioned in \cite{Knopki_2025}, but without linking it to the total charge. In \cite{Castro-Ruiz:2025yvi}, the authors present a scenario in which this ambiguity can be made explicit.
    
The momenta ambiguity does not appear in many previous works \cite{Loveridge:2017pcv, Carette:2023wpz, Bartlett:2006tzx, Glowacki:2023ooa} adopting the weak approach, where perspectival algebras are defined  as the image of a \textit{relativization} map. In the case of a tripartite system $ABC$, constrained by a finite symmetry group, an operator $T$ on the $BC$ subsystem in the perspective of $A$ would take the form 
\be \label{eq:yen}
\yen_A(T) = \sum_{g\in G} \ket{g}_A\bra{g} \otimes U^{BC}_g T U^{BC~ \dagger}_g 
\ee
where the $\yen$ map \cite{Bartlett:2006tzx, Carette:2023wpz, Castro-Ruiz:2025yvi} is defined as
\be
\yen: \textrm{Lin}(\ch_B \otimes \ch_C) \rightarrow \ca_{BC|A} \subset \ca_{\text{sym}}.
\ee
Importantly, this map is an injective homomorphism of C$^*$-algebras.

One can see that this choice of relativization map implies, in our previous translation group example, the selection of the subset of momenta given by  $\{P_1, P_2\}$, since these two operators are in the image of $\yen$. Furthermore, the proposed form of the $\yen$ map is motivated using the fact that it maps absolute positions to relative positions \cite{Carette:2023wpz}. However, one can reformulate the momenta ambiguity presented before precisely by saying that mapping absolute positions to relative positions is not sufficient to single out the $\yen$ map. Thus, the given form of the $\yen$ map seems to reflect a hidden physical assumption.

\subsection{Perspectives in the strong approach}
Choosing strong symmetry leads to a very different view of perspectives. 

As we already made explicit, imposing strong symmetry implies that $\ca_{\text{sym}} = \ca_{\text{S}}$ is a factor. As such it is isomorphic to perspectival algebras such as $A_{BC|A}$. Again, the point here is that in the weak case, the additional information that these algebras were missing compared to the symmetric algebra $\ca_{\text{sym}} = \ca_{\text{W}}$ was the total charge, and that by considering instead $\ca_{\text{sym}} = \ca_{\text{S}}$ we get rid of that information already at the level of the symmetric algebra, and thus encounter no problem when trying to write an isomorphism from it to the perspectival algebras.

Another way to state this in the context of the translation group example we exposed earlier is that the ambiguity we previously saw is now no more precisely since the charge is set to zero. Adopting the strong approach also provides a very natural way of isomorphically moving from a subsystem's perspective to another: one can simply compose the inverse isomorphism going from a perspective to the symmetric algebra with the isomorphism going from the symmetric algebra to the desired perspective, as is routinely done in papers following this approach. 

Furthermore, imposing strong symmetry also means that the ambiguity in the collaborative perspective, associated to $\ca_{\text{col}}$, vanishes. The different sets of momenta belonging to the perspectival algebras are now identical and thus they all give rise to the same internal algebra, containing the same information about the global charge, i.e.\ that it is null.

Because these transformations are isomorphic, the perspectival algebras are in fact all equal: they contain the same information, only labeled differently. Going back to the tripartite example from the previous part, this means that we have $\ca_{BC|A} =\ca_{AB|C} =\ca_{AC|B}$. In other words, with respect to our Definition \ref{def:persp},  these perspectives do not differ in terms of point 1 but only in terms of point 2. Note that perspectival algebras still differ at the level of single-system description, since for example we can have $\ca_{A|B} \neq \ca_{A|C}$.

\section{A review of arguments motivating each approach} \label{sec:lr}
In this section, we present and discuss arguments motivating or criticizing each approach. These arguments are scattered across the literature, and often formulated in vastly different languages. We translate them in common terms, namely the ones we have introduced earlier, to allow for easy comparison among them. This also provides a background our following discussion can refer to.

\subsection{Pro-strong approach arguments}
Let us start with several arguments that have been put forward to justify the choice of the strong approach.

The most prominent argument in favor of using strong symmetry is, as we already stressed, that doing so allows for invertible frame transformations. Choosing $\ca_{\text{sym}} = \ca_{\text{S}}$ allows to define invertible maps from the symmetric algebra to the perspectives, and thus one can easily define changes of perspectives by going from one perspective to the symmetric algebra, and then from the symmetric algebra to the new perspective. An explicit argument showing this can be found in \cite{Hoehn:2023ehz}, which we expose and translate into the language introduced in this paper in Appendix \ref{ann:transl}.
While this is a compelling argument, we believe that we should find out whether perspective transformations are reversible, rather than assuming that they should be. 

A second argument, found in the first section of \cite{Krumm:2020fws}, can be understood as arguing that the total charge is inaccessible. In this work, the authors describe the strong approach  as the one that ``\textit{disregards the external relatum altogether}''. To them, this ``external relatum'' seems to include the total charge, since it is projected out when imposing strong symmetry. Since the charge is a symmetric quantity,  this line of thought implies that that the set of internally accessible quantities is different than the set of symmetric quantities. We would then need a physical condition for a quantity to be accessible, which is not spelled out in this work. Another interpretation of their argument could be that the authors consider the charge to be accessible even in the constrained theory (so that the set of accessible quantities is the set of symmetric ones), but physically set to be zero. We note that perspectival algebras could not contain this information, but $\ca_{\text{col}}$ could. If this is the case, no physical justification is provided, nor is it clearly stated. Again, a physical justification of why that might be the case has never been made anywhere else, to the best of the authors' knowledge.

In \cite{Krumm:2020fws, Hoehn:2021flk}, the authors argue in favor of strong symmetry through a different line of logic. First, they extend the symmetry group from global translations, to relations-controlled translations, i.e.\ translations that are controlled on the  (group-valued) relations between the systems. Going back to Example \ref{ex:2}, these would be translations acting differently on states with different values of $r$.  This is a consequence of the set of three axioms they state all symmetry transformations should verify. The second axiom, in particular, singles out the zero-charge sector. However, we note that it could be slightly modified to treat all charge sectors on an equal footing, by requiring that, in a superposition of states with different relation vectors, symmetry transformations translate every state by the same amount. This extended symmetry already yields a smaller set of allowed operators than $\ca_{\text{W}}$, which are the operators in $\ca_{\text{W}}$ that do not allow coherence between relations within non-zero charge blocks. 

To further reduce the symmetric algebra to $\ca_{\text{S}}$, the authors then invoke another argument, which can be found in Lemma 7 of their work. It states that only mixed states in $\ca_{\text{S}}$ remain invariant under purification using an auxiliary system $A$, which does not need to be constrained by the symmetry. These states are then interpreted as ``\textit{invariant under a change of description, but also the quantum information that these states carry about other systems is invariant under a change of description''.} While this stands as a mathematical statement, we disagree with  its physical counterpart: it would appear more logical to ban these ``bad purifications'', rather than all the states to which they could potentially be applied. 

This purification argument has been extended in \cite{Mekonnen:2025znb} in the context of invariance under permutations. In this work, the notion of \textit{complete invariance} is introduced. This criterion  demands the invariance of systems under local symmetry transforms when composed. This leads to the restriction to a single (non necessarily zero) charge sector. This is completely valid in the context of this work, but we do not think similar arguments applied to other contexts would result in the same conclusion, nor escape the comments made above. Only an operational examination of the situation (which is done in \cite{Mekonnen:2025znb}), and not a mathematical statement made \textit{a priori} can help in the choice of symmetry criterion.

\subsection{Pro-weak approach arguments}
We now discuss here arguments in favor of imposing weak symmetry when selecting symmetric states.

The most prominent argument in favor of the weak approach lies in the fact that the total charge is, strictly speaking, symmetric: it is invariant under the group action. As such, there is no reason to discard it in the fashion of the strong approach. There is no need to include an external relatum to define it, and as a consequence, it should not be excluded from an external relata-free theory. This argument has been made several times before, e.g.\ in \cite{Castro-Ruiz:2021vnq, Castro-Ruiz:2025yvi}.

One can also motivate the weak approach by following a Bayesian argument, as done in \cite{Bartlett:2006tzx, Poulin:2005dn}. This stems from considering that states that are localized in the degree of freedom the group acts upon are ``states relative to a reference frame''. The objective being obtaining a theory free of external relata, this information is deemed superfluous. In analogy with the law of total probability, this leads to the weak approach as a result of averaging over all possible group elements (i.e.\ applying the weak twirl). About the charge, the fact that it is symmetric means that there is no need to have any knowledge \textit{a priori} about external quantities, such as any group-variant quantities. Therefore, the Bayesian prescription is to keep this information.

The weak approach has also been motivated for reasons pertaining to the mathematics of infinite dimension and non-compact groups. The mathematical tools used in the strong approach are not well defined for all groups $G$. For non-compact groups, the Hilbert space corresponding to the zero-charge sector might simply not exist (see \cite{Carette:2023wpz} and references therein for more details). There are two objections we can make about this argument. First, when faced with a well-defined problem using a fixed group, like we have done for $\Z_n$, it seems unsatisfactory to argue against using a framework solely because it might be mathematically ill-defined in the case of different groups. Second, this problem appears to be a mathematical technicality which could be solved using techniques from the theory of distributions. The formal framework used to this end is called refined algebraic quantization, for more details see \cite{delaHamette:2021oex} and references therein. Ideally, we would want to discriminate between the strong and weak approaches based on physically motivated arguments, and then formulate a mathematical framework based on those.

\section{An operational proposal and how it leads to the weak approach} \label{sec:oa}

The previous section paints a simple picture. Situations where the charge associated to the symmetry is deemed nonphysical, or, with equivalent consequences, inaccessible to internal observers (i.e.\ not in $\ca_{\text{col}}$) call for the use of strong symmetry. On the other hand, we should use weak symmetry in situations in which the charge is in the collaborative algebra $\ca_{\text{col}}$.
It thus all comes down to a basic physical question: is the total charge accessible?
In this section, we consider this problem in the context of the simplest possible scenario, involving two systems. We propose two postulates pertaining to the physical abilities of our two agents, and show how they imply that the answer to the previous question is positive.

\subsection{Motivation}

Before diving into the toy model we want to introduce, let us take a step back to specify what we are trying to achieve. Up until this moment, the point we have made our best effort to highlight in the present paper is that ``this given system is constrained by a symmetry'' is an ambiguous statement. As it stands, this statement can lead to two physically different situations: the system is either strongly symmetric or weakly symmetric. This, as we have showed, is equivalent to the accessibility, or lack thereof, of the total charge. The dichotomy is clear, and does not pose any sort of problem on its own. The difficulty resides in deciding this accessibility puzzle.

The answer of course lies in the study of the physical situation one is faced with. What does being symmetric mean in this very situation? Which degrees of freedom are concerned? Answering these questions should stem from the postulation of operational capacities for the observers involved, and from those, one can derive a stance on the accessibility of the total charge. Indeed, the most satisfactory answer to the question of total charge accessibility is the demonstration of an operational procedure showcasing how precisely it would be accessed, or clear operational reasons why it is inaccessible. Both options can be worked out starting from the operational capabilities the agents acting on the system are assumed to have, which need to be clearly made explicit. An instance of such a setting is what we aim at describing in the rest of this paper.

\subsection{Goal and strategy} \label{sec:g&s}
Our goal here is to propose a physically motivated solution to the question: what is the content of the collaborative algebra $\ca_{\text{col}}$? As highlighted in the previous sections, the crucial task at hand is  to find out whether the total charge observable is included in it. Remembering the definition of the internal algebra as the algebraic span of the perspectival algebras, we can reduce our problem to the following question: what do the perspectival algebras contain? For simplicity, we restrict to  a scenario with only two systems, corresponding to agents which we name Alice and Bob (see our discussion about agents in Section \ref{sec:agents}), where the question becomes: what is inside of $\ca_{B|A}$ (of course, the situation for $\ca_{A|B}$ is then symmetric)?  As we made clear in Section \ref{sec:cons}, part of the content of the perspectival algebras has been consensual from the start: the relative positions\footnote{We refer to the position as the quantity the symmetry group acts on.} are in $\ca_{B|A}$.  The question is about which momenta\footnote{By this we mean the quantity conjugate to the position, in analogy with the mechanical meaning of the term.} the perspectival algebras contain, precisely since the commutation relations with the position are not restrictive enough to single out a unique choice of associated momenta. We thus see that the crucial question for us, from which the rest will derive, is: which physical quantity Eve (the observer associated with the kinematical algebra) describes as being measured by Alice when Alice measures Bob's momentum. This cannot be derived from a mathematical formalism, since which mathematical formalism to use is what we are trying to uncover. We want to go the other way: argue in favor of a mathematical formalism by saying it fits the operational procedure by which Alice measures relative quantities of Bob.

Our strategy to reason operationally is the following: we want to 1/ describe an experimental procedure that Alice can perform to find out Bob's momentum relative to her (and symmetrically for Bob); and 2/ understand what this procedure looks like in Eve's description.
Necessarily, this will be based on a number of fundamental postulates, in particular about which procedure is relevant here, as well as how it is described formally. We present them for the sake of the clarity of our argument's foundations, and explicitly refer to them when we rely on them.

Our first postulate is that Alice can measure Bob's momentum by performing an interference experiment on Bob's degree of freedom. The second postulate describes what this procedure looks like in Eve's description. Using these two postulates, we are able to state what Eve says Alice has measured. In other words, we are able to infer what the momentum Alice attributes to Bob is in Eve's description. To the best of the authors' knowledge, such an argument has never been made in the literature before. As we showed in Section \ref{sec:lr}, past efforts in the literature have focused on extending and applying frameworks in which  the content of the collaborative algebra $\ca_{\text{col}}$ is derived from the mathematical formalism, but not operationally justified.

\subsection{Scenario : Setting the scene} \label{sec:secene}

\begin{figure}
    \centering
    \includegraphics[width=0.7\linewidth]{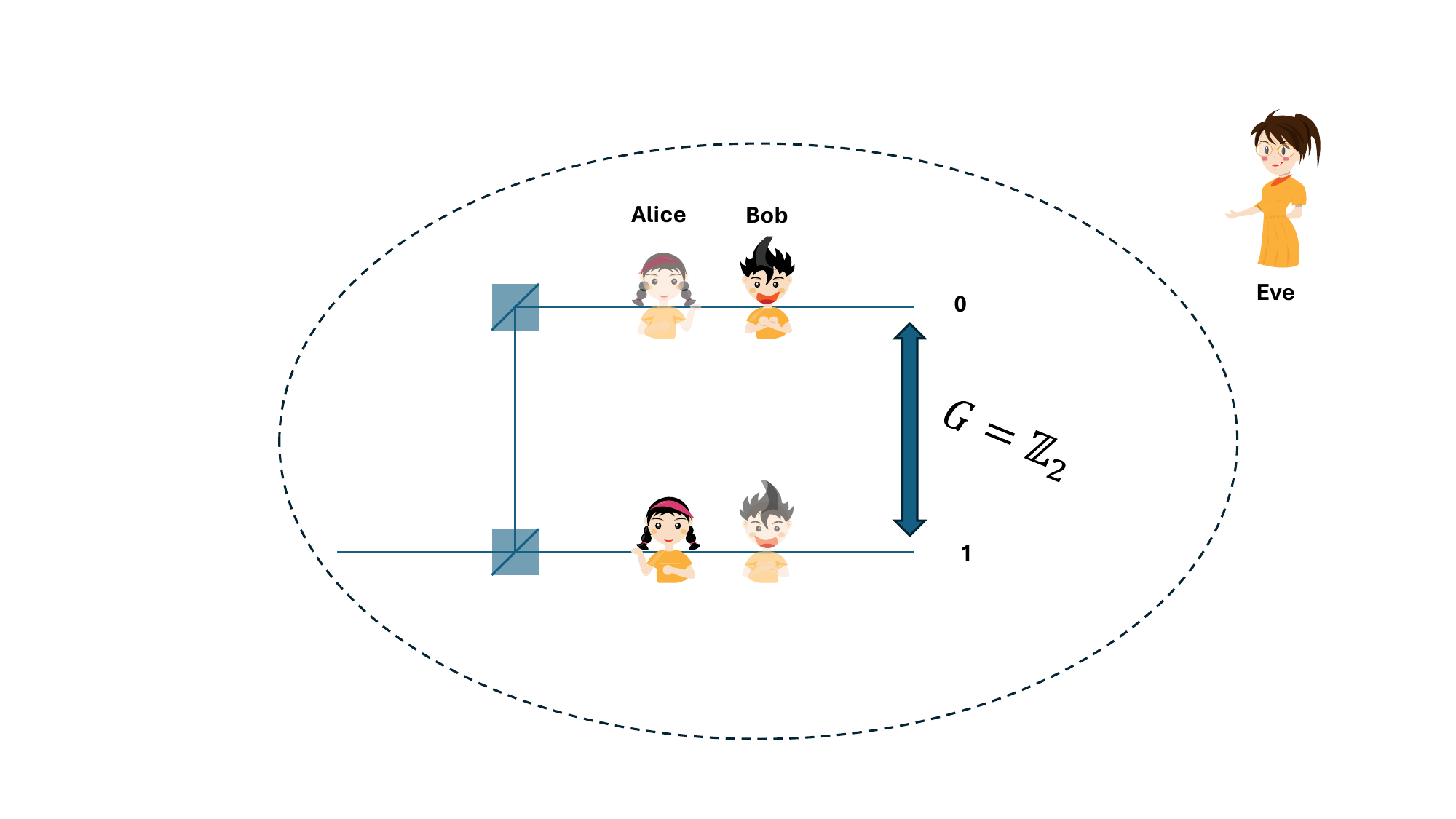}
    \caption{\em Experimental set-up from Eve's perspective}
    \label{fig:app}
\end{figure}

Let us now turn to the the concrete application of our strategy. We start by setting the scene. We consider a physical situation, consisting in the propagation of physical systems through two paths. This set-up is informally pictured in Figure \ref{fig:app}.  This scenario enjoys a very simple mathematical  setting; in particular, the symmetry group is finite and abelian. We believe that starting with such little technicality will allow for a clear conceptual understanding, and we leave the generalization to more elaborate cases to future work.

\begin{figure}
    \centering
    \includegraphics[width=0.9\linewidth]{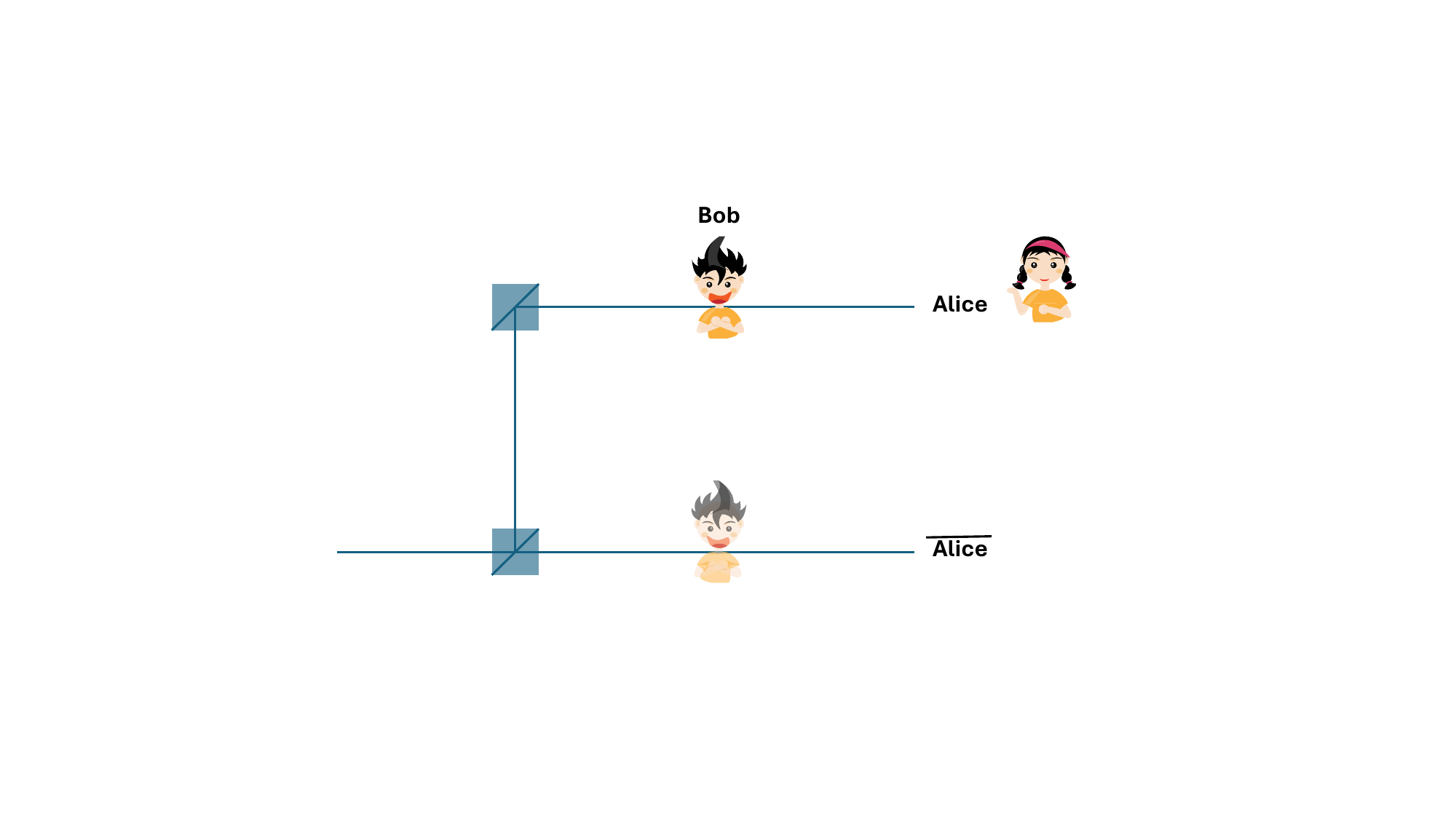}
    \caption{\em Experimental set-up from Alice's perspective}
    \label{fig:as}
\end{figure}

This set-up can be described from different perspectives. We first introduce Eve, the super observer, unconstrained by symmetries and having access to the full kinematical algebra $\ca_{\text{kin}}$. Eve labels one path $0$, and  the other path $1$. Next, to each of the two systems in the apparatus, we associate an agent, Alice and Bob.\footnote{See our discussion in Section \ref{sec:agents} for what in means to associate a laboratory or an agent to a system.  As a reminder, we can define the perspective related to a given system as the operations accessible to an agent restricting themselves to act only on the degrees of freedom relative to the system.} Each of Alice and Bob only describes a single degree of freedom in the present situation: which path of the apparatus the other  system travels through relative to their own. The situation in Alice's perspective is informally pictured in Figure \ref{fig:as}.

Formally, Eve assigns to the systems a Hilbert space $\ch_{\text{kin}} = \ch_{A|E} \otimes \ch_{B|E}$ where each tensor factor is spanned by $\{ \ket{0}_I, \ket{1}_I \}$ where $I = A|E,B|E$. For example, in the case where both Alice and Bob are sent through the $0$ path, Eve describes the system as being in the state $\ket{0}_{A|E} \ket{0}_{B|E}$. 
Eve can also describe the situation using a formulation in terms of modes, which we will use a lot and will turn out to be crucial. By this we mean that she can describe the physical situation in terms of occupation of the paths of the apparatus, which could then be either empty, contain Alice, Bob, or both, which one can think of as excitations of the corresponding modes. In accordance with the set-up's specifications, we want to keep one of each system A and B, and therefore forbid some combinations of the aforementioned modes to be populated. Mathematically, this can be described using another tensor factorization, inside which we only work with states matching the needs of our physical situation.

In our case, there are four possible modes, representing each a difference piece of information: whether Alice is in the $0$ path ($\ch_{A,0}$), whether Alice  is in the $1$ path ($\ch_{A,1}$), whether Bob  is in the $0$ path ($\ch_{B,0}$), and whether Bob is in the $1$ path ($\ch_{B,1}$). Each of these mode can be either empty ($\ket{\textrm{vac}}$), or occupied ($\ket{\textrm{occ}}$). Tensored together, these modes form a space that is larger than $\ch_{\text{kin}}$, but some of its subspaces are not relevant in the present situation, and should therefore be excluded from the description.\footnote{We stress that these states are excluded because they do not fit the scenario we are describing, and not because they would not satisfy some symmetry condition.} More precisely, in the scenario at hand there always are two excitations, one for Alice and one for Bob, so we can restrict our analysis to the space spanned by the family 
\be \label{eq:sc}
\begin{split}
\{ \ket{\textrm{occ}}_{A, 0} \ket{\textrm{vac}}_{A, 1} \ket{\textrm{occ}}_{B, 0} \ket{\textrm{vac}}_{B, 1},\\ 
\ket{\textrm{occ}}_{A, 0} \ket{\textrm{vac}}_{A, 1} \ket{\textrm{vac}}_{B, 0} \ket{\textrm{occ}}_{B, 1}, \\
\ket{\textrm{vac}}_{A, 0} \ket{\textrm{occ}}_{A, 1} \ket{\textrm{occ}}_{B, 0} \ket{\textrm{vac}}_{B, 1},\\
\ket{\textrm{vac}}_{A, 0} \ket{\textrm{occ}}_{A, 1} \ket{\textrm{vac}}_{B, 0} \ket{\textrm{occ}}_{B, 1} \}.
\end{split}
\ee
This restriction corresponds to selecting the states matching our physical situation.
To ensure the conservation of this selection, we ban any mapping from the space spanned by the family above to any other part of $\ch_{A,0}\otimes \ch_{A,1}\otimes \ch_{B,0}\otimes \ch_{B,1}$. In other words, this means enforcing that the unitary maps describing the system's evolution are null on and to the orthogonal complement to \ref{eq:sc}.
The relation between the description in terms of systems and the one in terms of modes is given by the isomorphism between $\ch_{\text{kin}}$ and the space spanned by (\ref{eq:sc}), which reads
\be
\begin{cases}
    \ket{0}_{A|E}\ket{0}_{B|E} &\mapsto  \ket{\textrm{occ}}_{A, 0} \ket{\textrm{vac}}_{A, 1} \ket{\textrm{occ}}_{B, 0} \ket{\textrm{vac}}_{B, 1} \\
    \ket{0}_{A|E}\ket{1}_{B|E} &\mapsto \ket{\textrm{occ}}_{A, 0} \ket{\textrm{vac}}_{A, 1} \ket{\textrm{vac}}_{B, 0} \ket{\textrm{occ}}_{B, 1} \\
    \ket{1}_{A|E}\ket{0}_{B|E} &\mapsto \ket{\textrm{vac}}_{A, 0} \ket{\textrm{occ}}_{A, 1} \ket{\textrm{occ}}_{B, 0} \ket{\textrm{vac}}_{B, 1} \\ 
    \ket{1}_{A|E}\ket{1}_{B|E} &\mapsto \ket{\textrm{vac}}_{A, 0} \ket{\textrm{occ}}_{A, 1} \ket{\textrm{vac}}_{B, 0} \ket{\textrm{occ}}_{B, 1} .
\end{cases}
\ee

\subsection{Operations inside and out}
Next, let us see which operations Alice and Bob themselves are able to perform. In order to open the door for quantum reference frames, we suppose that the following symmetry holds: the absolute position of Alice and Bob (i.e.\ which path they travel in) carries no physical meaning to these agents. This symmetry is described by the group $\Z_2$ acting on the Hilbert space $\ch_{\text{kin}}$ via the unitary representation $\{ U_0 = \mathbb{1}, U_1 = X_{A|E} \otimes X_{B|E}  \}$, where $X_I \ket{0}_I = \ket{1}_I, X_I \ket{1}_I = \ket{0}_I$.

 As we explained, each of Alice's and Bob's perspective is associated to an algebra, respectively $\ca_{B|A}$ and $\ca_{A|B}$. In terms of Hilbert spaces, the algebra associated to Alice's perspective, $\ca_{B|A}$, is isomorphic to the algebra of operators over a Hilbert space denoted $\ch_{B|A}$. As we mentioned, this space is essentially a qubit: $\ch_{B|A} =\text{Span} \{ \ket{0}_{B|A}, \ket{1}_{B|A} \}  \cong \C^2$. Alice attributes to Bob the state $\ket{0}_{B|A}$ if he is in the same path as her, and the state $\ket{1}_{B|A}$ if he is in the opposite path.  
Like we did for Eve's description, we introduce a description in terms of modes, each mode corresponding to the presence of Bob's system in each path. To this end, we will work with another larger Hilbert space, which can be decomposed as $\ch_{A} \otimes \ch_{\bar A}$. Each of these tensor factors corresponds to a qubit, indicating whether the corresponding path is occupied by Bob in Alice's perspective. The subscript $A$ indicates Alice's path, and $\bar A$ the one she is not on. Each of these modes is spanned by $\{ \ket{\textrm{occ}}_J, \ket{\textrm{vac}}_J \}_{J=A, \bar A}$. This space is larger than $\ch_{B|A}$, but the latter is isomorphic to $\text{Span} \{ \ket{\textrm{occ}}_A \ket{\textrm{vac}}_{\bar A},  \ket{\textrm{vac}}_{A} \ket{\textrm{occ}}_{\bar A}\}$. The natural way to write out the isomorphism connecting them is 
\be \label{eq:iso}
\begin{cases} 
    \ket{0}_{B|A} &\mapsto \ket{\textrm{occ}}_A \ket{\textrm{vac}}_{\bar A} \\
    \ket{1}_{B|A} &\mapsto \ket{\textrm{vac}}_A \ket{\textrm{occ}}_{\bar A}\,.\\
\end{cases}
\ee
Here again, this is simply selecting the states matching our situation, in an analogous fashion to what we previously presented.

As we exposed in Section \ref{sec:g&s}, we are interested in understanding which observables are in $\ca_{B|A}$. Importantly, we want to do so by conducting an analysis of the relevant operational procedure for measuring these observables, which we now present in detail.

\subsection{Measurements in general and delocalized registers}

We are going to describe all the experimental procedures from the point of view of Alice, and then from the point of view of Eve. This is an important first step before presenting the measurement of specific quantities like position. Here, we describe how a general measurement performed by Alice is depicted. This way, we are able to focus on the assumptions we make, and leave the implications as to which observable is measured to a later mathematical proof deriving from these assumptions.

We assume that Alice performs a measurement by accessing and writing in a physical system acting as a register that is local in her perspective (in which Alice herself is indeed always localized). When we say that Alice has access to a certain quantity, or can measure it, we mean that Alice can collect the value of this quantity and store it in her local register. Should Alice be delocalized with respect to Eve, so would this register be.  In this case, in Eve's description,  Alice's measurement appears as two different interactions, leading to the value measured being stored in a superposition of two different registers. Note, however, that this does not mean that Alice's measurement result is inaccessible to Eve. Eve describes these two registers as two modes, and, by having her own register interact, in a controlled manner, with a mode then the other, she can retrieve the value written in Alice's register. Importantly, it is easy to see that she can do this without copying (and therefore decohering) Alice's position. 

Leveraging the partition into modes that we introduced earlier, we can depict the situation using diagrams. In Alice's perspective, a diagram has three wires, each corresponding to a Hilbert space isomorphic to a qubit. We have one wire for each path of the apparatus labeled $A$ and $\bar A$, as well as a wire for Alice's local register, which we represent as a dashed wire next to Alice's. 
\be
\begin{quantikz}[wire types = {n,q,q}]
 \lstick{$\ket{0}$} &&  &&  \rstick{A's local register} \\
&&  && \rstick{$A$} \\[1cm]
&& && \rstick{$\bar A$}
\arrow[from=1-1,to=1-5,black,dashed,-]{}
\end{quantikz}
\ee
On the other hand, in Eve's description, we have six modes, divided in two groups, one for each path of the apparatus, labeled $0$ and $1$. In each path, we have a wire for Alice, a wire for Bob, as well as a wire for each of the two local modes that (from Eve's point of view) make up Alice's register:
\be
\begin{quantikz}[wire types = {n,q,q,n,q,q}]
 \lstick{$\ket{0}$} &&&  &&&  \rstick{A's local register, $0$} \\
&&&  &&&  \rstick{$A, 0$}  \\
 &&&  &&&  \rstick{$B, 0$}\\ [1cm]
\lstick{$\ket{0}$} &&&  &&&  \rstick{$A$'s local register, $ 1$} \\
 &&&  &&&  \rstick{$A, 1$}  \\
 &&&  &&&  \rstick{$B, 1$} \\
 \arrow[from=1-1,to=1-7,black,dashed,-]{}
\arrow[from=4-1,to=4-7,black,dashed,-]{}
\end{quantikz}
\ee

Let us look at how a measurement interaction $U$ performed by Alice would look like. Alice performs this operation on both B and her local register. We look at the case where Alice only performs the operation $U$ on Bob if he is on her path, which yields the circuit
\be
\begin{quantikz}[wire types = {n,q,q}]
 \lstick{$\ket{0}$} &&\gate[2]{U}&&  \rstick{$A$'s local register} \\
 &&& & \rstick{$A$} \\[1cm]
 && &&\rstick{$\bar A$}
 \arrow[from=1-1,to=1-5,black,dashed,-]{}
\end{quantikz}
\ee
In Eve's description, the same operation appears controlled on Alice's position, and the measurement result is only saved to the mode corresponding to Alice's register on the corresponding path. If Alice's position is in a superposition in Eve's description, then so is the register and thus the measurement result. As a circuit, this is represented as
\be \label{eq:eve_pov}
\begin{quantikz}[wire types = {n,q,q,n,q,q}]
 \lstick{$\ket{0}$} && \gate[2]{U}& && \rstick{$A$'s local register, $0$} \\
 &&& &&  \rstick{$B,0$} \\
&& \ctrl{-1} &  && \rstick{$A,0$}\\ [1cm]
\lstick{$\ket{0}$} &&\gate[2]{U}&  && \rstick{$A$'s local register, $1$} \\
  && &  &&  \rstick{$B,1$}  \\
&& \ctrl{-1} &  && \rstick{$A,1$}\\
\arrow[from=1-1,to=1-6,black,dashed,-]{}
\arrow[from=4-1,to=4-6,black,dashed,-]{}
\end{quantikz}
\ee

\subsection{Relative position measurement} \label{sec:posm}

We are now well equipped to understand how Alice and Bob would measure each other's position, and what this looks like from the point of view of Eve. To measure Bob's position with respect to herself, Alice uses a simple controlled operation: she writes $ 1$ in her local register if Bob is in the same path of the apparatus as her, and $0$ if he is not. This operation is described from the point of view of Alice by\footnote{The controlled operation is applied when the control is in the $\ket{\textrm{occ}}$ state.}

\be
\begin{quantikz}[wire types = {n,q,q}]
 \lstick{$\ket{0}$} && \targ{} &&  \rstick{$A$'s local register} \\
 && \ctrl{-1} && \rstick{$A$} \\[1cm]
 && && \rstick{$\bar A$}
 \arrow[from=1-1,to=1-5,black,dashed,-]{}
\end{quantikz}
\ee

To understand what this operation looks like in Eve's description, we can now simply apply the rule described in (\ref{eq:eve_pov}), yielding the circuit

\be
\begin{quantikz}[wire types = {n,q,q,n,q,q}]
 \lstick{$\ket{0}$} &&& \targ{} &&&  \rstick{$A$'s local register, $0$} \\
 &&& \ctrl{-1 } &&&  \rstick{$A,0$} \\
&&& \ctrl{-2 } &&&  \rstick{$B,0$}\\ [1cm]
\lstick{$\ket{0}$} &&& \targ{} &&&  \rstick{$A$'s local register, $1$} \\
  &&& \ctrl{-1 } &&&  \rstick{$A,1$}  \\
&&& \ctrl{-2 } &&&  \rstick{$B,1$}\\
\arrow[from=1-1,to=1-7,black,dashed,-]{}
\arrow[from=4-1,to=4-7,black,dashed,-]{}
\end{quantikz}
\ee
As it stands, this is itself a postulate, but since there is nothing controversial about it (everyone would agree that relative position observables are in the perspectival algebras) we do not label it as such. This operation yields what we intuitively expected: if Alice sees Bob in the same branch as her, she writes 1 in her register, and if she does not, she writes 0.
Bob's measurement of Alice's position relative to him follows analogously. 

\subsection{Momentum measurements} 
Bearing this in mind, let's see how this can help in understanding momentum measurements. Momentum measurements are interference experiments. In order to describe an interference experiment, we first need to introduce ``beam splitters''.\footnote{We use here the term beam splitter as an analogy to the optical case. In our situation, we describe an operation which acts as a beam splitter would on an optical system.} By a beam splitter -- defined with respect to a pair of maximally incompatible observables -- we mean a unitary that, when given an eigenstate of one observable, returns an eigenstate of the other. In a way, it ``localizes'' the first observable with respect to the second. Therefore, in our case, a beam splitter is a device that acts as the following unitary map on a two-qubit state in the description in terms of modes:
\be \label{eq:bs}
\begin{cases}
    \ket{\textrm{occ}}\ket{\textrm{vac}}&\longleftrightarrow \frac{1}{\sqrt{2}} (  \ket{\textrm{occ}}\ket{\textrm{vac}}+\ket{\textrm{vac}}\ket{\textrm{occ}} ) \\
    \ket{\textrm{vac}}\ket{\textrm{occ}}&\longleftrightarrow \frac{1}{\sqrt{2}} (\ket{\textrm{occ}}\ket{\textrm{vac}} - \ket{\textrm{vac}}\ket{\textrm{occ}}) \\
\end{cases}
\ee
In light of the isomorphism in (\ref{eq:iso}), we can see that, when described in terms of positions, a beam splitter is simply a Hadamard gate.
Importantly, because this is a two-qubit gate, it is crucial to distinguish the first qubit from the second in the perspective one is interested in. Note that Alice can of course make this distinction without breaking the $\Z_2$ symmetry she is subjected to in Eve's perspective. She simply labels the two branches of the apparatus, and this labeling is completely unrelated to the labeling Eve does on her side. This choice that Alice makes is a convention, which is going to be important in postulating how Eve describes a beam splitter Alice applies on Bob.

\begin{figure}
    \centering
    \includegraphics[width=0.7\linewidth]{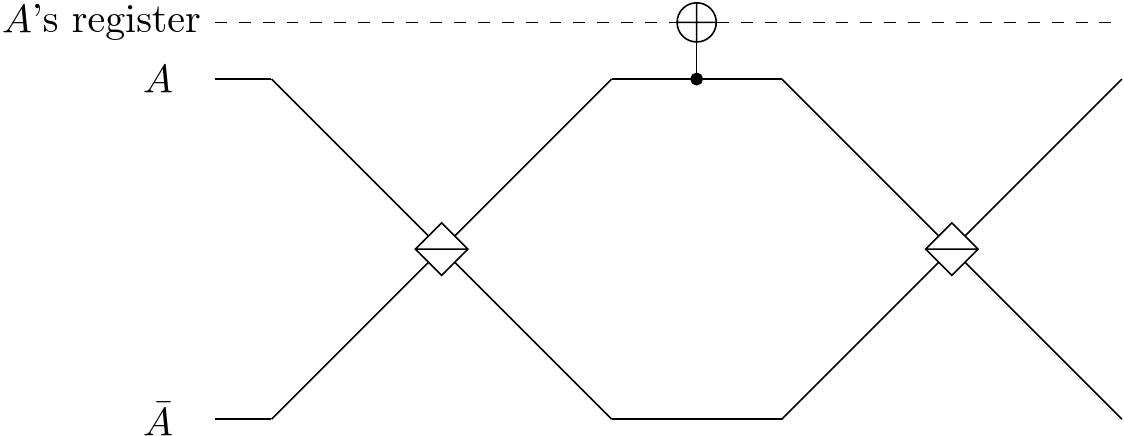}
    \caption{\em Postulate 1: Alice's measurement of Bob's momentum in her own perspective}
    \label{fig:A's momentum}
\end{figure}

The next step in our development is to understand how Alice would measure Bob's pseudo momentum relative to her, and what this looks like both  in Alice's perspective and in Eve's. Echoing our discussion in Section \ref{sec:secene}, this is precisely where where we introduce crucial postulates on which our conclusion relies. The first one concerns the form of the measurement of Bob's momentum by Alice in her own description. 
As we highlighted in Section \ref{sec:persp}, this will exhaust the content of Alice's perspective on Bob $\ca_{A|B}$. Indeed, we showed in Section \ref{sec:posm} how Alice can access Bob's position $X_{A|B} \in \ca_{A|B}$. What is left to understand is what is the conjugate pseudo-momenta also contained in this perspective.
This postulate solely concerns Alice's description, and not others'. 
\begin{postulate}
  To measure Bob's momentum, Alice feeds the two modes (the one on her side and the one on the other) into the beam splitter.  She then says that she measures Bob's momentum in the $+$ state if Bob's system ends up in the path Alice is in after the beam splitter.\footnote{This choice between + and - is a convention, corresponding to the description of the beam splitter we chose to give. It does not affect our conclusion.} As a circuit, this is depicted in Figure \ref{fig:A's momentum}.
\end{postulate}

Let us go through a simple example to illustrate how this measurement works. 
\begin{example}
Suppose for instance that, in Alice's perspective, Bob is in a superposition of being in the same path as Alice and being in the opposite path. Alice wants to work out the charge (either + or -) between the  branches of the superposition, or in other terms, discriminate between 
\be
\begin{cases}
    \ket{+}_{B|A} &= \frac{1}{\sqrt{2}} (\ket{\textrm{vac}}_A \ket{\textrm{occ}}_{\bar A} + \ket{\textrm{occ}}_A \ket{\textrm{vac}}_{\bar A}  ) \\
    \ket{-}_{B|A} &= \frac{1}{\sqrt{2}} (\ket{\textrm{occ}}_A \ket{\textrm{vac}}_{\bar A}-\ket{\textrm{vac}}_A \ket{\textrm{occ}}_{\bar A} ). \\
\end{cases}
\ee
To this end, she sends the state she is given into the apparatus of Figure \ref{fig:A's momentum}, where it reaches the beam splitter. Using (\ref{eq:bs}) we see that the $A$ path on the diagram will contain $\ket{\textrm{occ}}_A$ if and only if the state $\ket{+}$ was passed. Therefore, Alice's register (being initialized in the $\ket{0}$ state) will contain the value $1$ if and only if Alice measures Bob in her path after the beam splitter. 
\end{example}
In a nutshell, in terms of the factorization into modes, measuring the momentum is simply conjugating a position measurement between two beam splitters.

\begin{figure}
\centering
   \begin{subfigure}[h]{8cm}
   \centering
    \includegraphics[width=6.98cm, height=3.9cm]{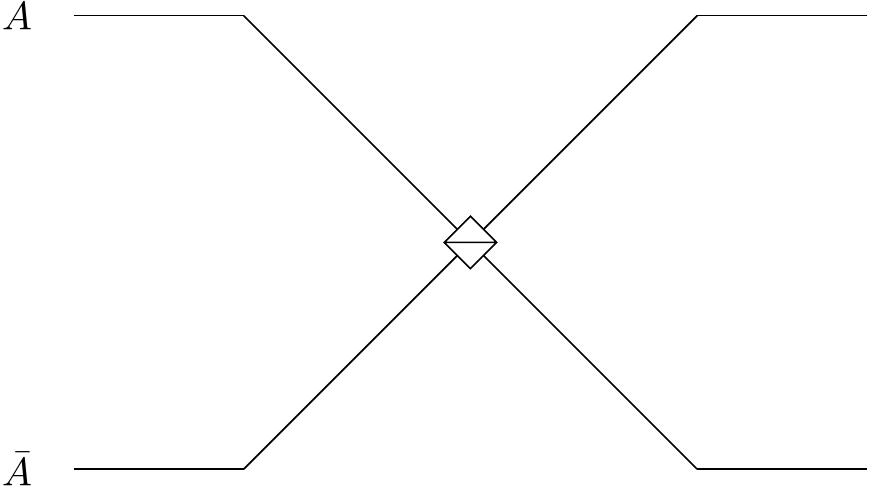}
    \caption{\em Alice's description}
    \label{fig:A_bs}
    \end{subfigure}
    ~~~
    \begin{subfigure}[h]{8cm}
    \centering
    \includegraphics[width=7.76cm, height=6.08cm]{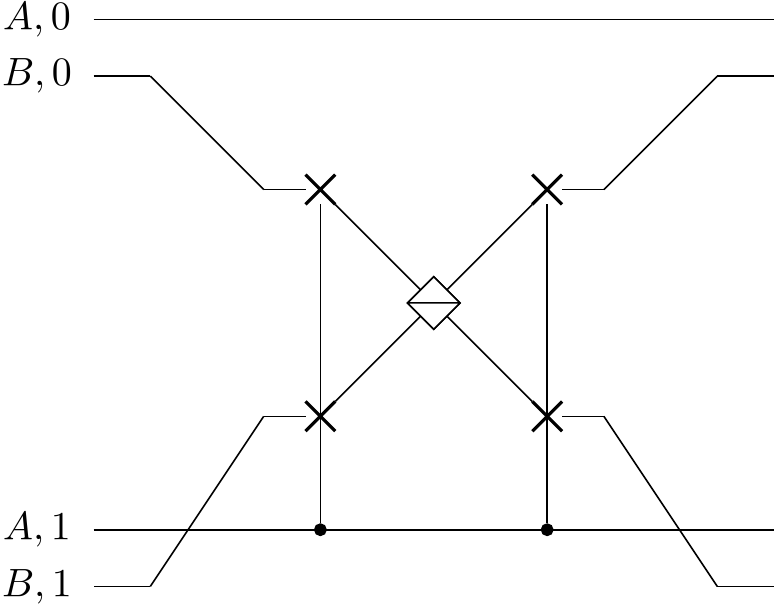}
    \caption{\em Postulate 2 : Eve's description}
    \label{fig:E_bs}
    \end{subfigure}

    \caption{\em Alice performing a beam splitter on Bob's position as seen by different agents.}
    \label{fig:bs}
\end{figure}

\begin{figure}
    \centering
    \includegraphics[width=\linewidth]{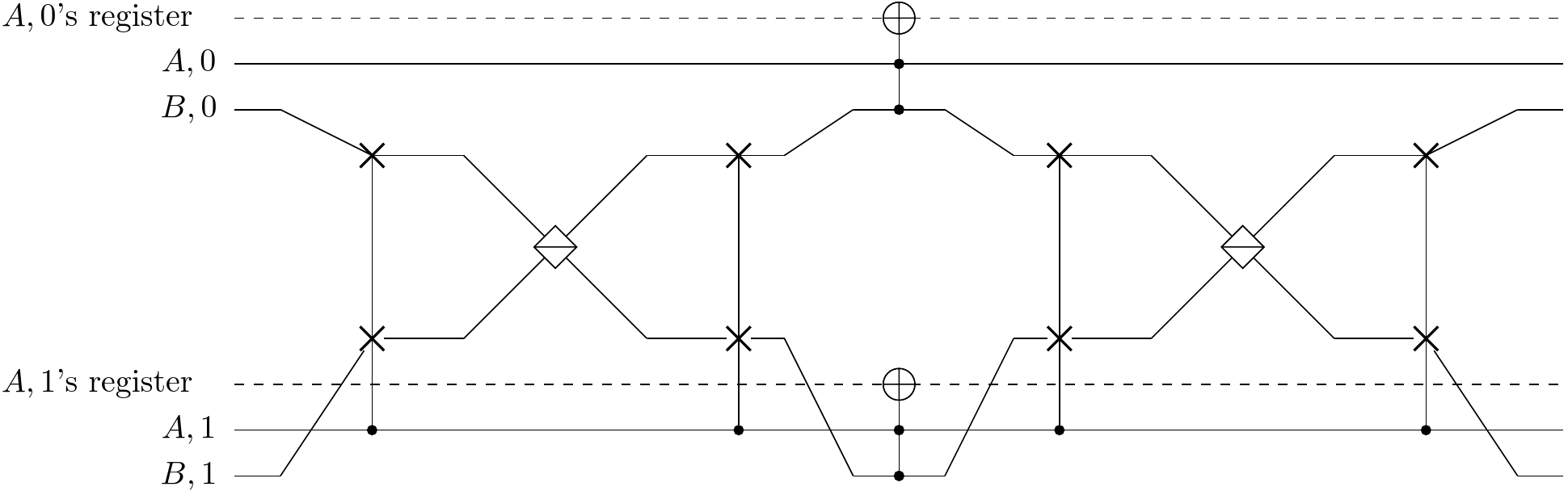}
    \caption{\em Alice's measurement of Bob's momentum in Eve's description}
    \label{fig:E_momentum}
    \end{figure}

Let us now discuss how Eve would describe this measurement of Bob's momentum by Alice. Since we already know what a position measurement looks like from the point of view of Eve, the only remaining question is: what does a beam splitter look like from Eve's description? 
\begin{postulate}
    Eve's description of the application of a beam splitter by Alice is given by Figure \ref{fig:E_bs}. As a direct consequence of this statement and Postulate 1, Eve's description of the measurement of Bob's momentum by Alice is given in Figure \ref{fig:E_momentum}.
\end{postulate}

This is simply a beam-splitter sandwiched between two controlled SWAPs. The guiding insight here is that the orientation of the beam splitter (i.e.\ which path is up or down, or $0$ or $1$) can be different for Eve and Alice. Eve has to account for this by adapting her description according to Alice's position via coherent control. This means swapping Bob's position controlled on Alice's position, as is depicted on Figure \ref{fig:E_bs}.

We now have everything we need to obtain Eve's description on Alice's measurement of Bob's momentum. Combining them results in the measurement process pictured in Figure \ref{fig:E_momentum}.
\begin{example}
    Eve initializes the circuit in the state
    \be
    \ket{\psi} = \ket{\textrm{occ}}_{A,0} \ket{\textrm{vac}}_{A,1} \left ( \frac{\ket{\textrm{vac}}_{B, 0} \ket{\textrm{occ}}_{B, 1} + \ket{\textrm{occ}}_{B, 0} \ket{\textrm{vac}}_{B, 1}}{\sqrt{2}} \right ) \equiv \ket{\textrm{occ}}_{A,0} \ket{\textrm{vac}}_{A,1} \ket{+}_{B|E} \, .
    \ee
    Let us follow this state as it propagates through the circuit. First, it reaches the controlled-swap. This operation is controlled by the state of $A,1$, which is currently $\ket{\textrm{vac}}$, so it acts as the identity. Then comes the beam splitter on the two $B$ modes. According to (\ref{eq:bs}), we have 
    \be
    \ket{\textrm{occ}}_{A,0} \ket{\textrm{vac}}_{A,1} \ket{+}_{B|E} \xlongrightarrow{B.S.} \ket{\textrm{occ}}_{A,0} \ket{\textrm{vac}}_{A,1} \ket{\textrm{occ}}_{B,0} \ket{\textrm{vac}}_{B,1} \, .
    \ee
    After that comes another idled swap, then the CNOT on the registers. The latter are both initialized in the $\ket{\textrm{vac}}$ state. In this case, the CNOT gate in the $0$ path gets activated, while the other one does not, as clearly shown by the previous equation. Therefore, the register in the $0$ path, where Alice is, ends up being in the state $\ket{\textrm{occ}}$ corresponding to measuring charge $+$.  Thus, in this example, Alice and Eve agree on what they call Bob's momentum.
\end{example}

We are now able to state the main result of this section, answering the question we formulated earlier.
\begin{theorem} \label{main point}
    Given Postulates 1 and 2, the momentum Alice attributes to Bob $X_{B|A}$ and the momentum Eve measures $X_{B|E}$ are the same:
    \be
    X_{B|A} = X_{B|E}.
    \ee    
\end{theorem}

The proof of this theorem is given in Appendix \ref{ann:mainpoint}. Theorem \ref{main point} answers the question we set as our main interrogation in Section \ref{sec:g&s}. Alice's perspectival algebra contains $X_{B|A}$, and this is precisely also the momentum Eve describes as being Alice's measurement of Bob's momentum. As a direct consequence, we also have $X_{A|B} = X_{A|E}$. These facts directly answer our original question about the content of $\ca_{\text{col}}$, through 
\be 
  X_{A|E} X_{B|E} = X_{A|B} X_{B|A} \in \ca_{A|B} \vee \ca_{B|A} = \ca_{\text{col}}.
\ee
The total charge is in $\ca_{\text{col}}$. This quantity is relevant to symmetry-constrained observers, and should therefore not be discarded when seeking a formalism describing their perspectives. Therefore, our analysis leads to favoring the weak approach.

In the end, this simple example gives an instance of the type of argument we were looking for. It manages to motivate the weak approach over the strong one on operational terms. The mathematical assumptions that were used to formulate it were consequences of operational postulates, insisting instead on the conceptual aspects of the proposal. Furthermore, we did this while validating a specific choice of momenta in each perspective, also solving the momenta ambiguity exposed in Section \ref{eq:weakpersp}.

\section{Conclusion}

This paper made two main points. First, the main physical difference between QRF frameworks -- mathematically embodied by the different group-averaging methods they employ -- is total charge accessibility. This point brings conceptual clarity to the differences between the frameworks, which we hope can help in future work involving them.

Second, we proposed an operational argument to discriminate between the two stances on charge accessibility. We formulated reasonable assumptions about the operational abilities of symmetry-constrained agents taking part in an idealized scenario, using the very user-friendly symmetry group $\Z_2$. From these postulates, we deduced that the agents can collaborate to determine the value of the total charge, making us conclude in favor of the weak approach. Again, we stress that this conclusion entirely depends on the postulates we formulate. We would gladly welcome substantiated challenges to their physical naturalness, and a general debate around the operational capacities of symmetry-constrained observers.

In the process of exposing these main points, we produced a few other noteworthy results. First, we formulated a definition of a perspective based on physical restrictions, which we used in our operational argument. Second, by formulating the differences between approaches in the language of C$^*$-algebras , we made them more explicit, and we used this structural tool to explain the non-invertibility of perspective transformation in the weak approach. Third, in our operational example, by understanding the relation between the algebra of momenta in Eve's and Alice's perspectives, we were able to lift the ambiguity in momenta we unearthed when discussing perspectives in the weak approach. Finally, in doing so, we also highlighted how  a register localized in one perspective could be delocalized in another. We leveraged this fact to derive our conclusion by remarking that this register can be probed without decohering the other degrees of freedom associated to the corresponding agent.

Our work has an obvious limitation: the fairly simple type of symmetry structure to which we restricted both to make our point about charge accessibility and to construct our operational toy model. In fact, we purposefully looked to strip our operational argument down to the simplest possible mathematical context. Indeed, we would consider it highly unlikely that the core of the underlying physics governing which quantities are available to symmetry-constrained observers would hinge on the nature of the group involved, since the problem already presents itself in our $\Z_2$ example. Nevertheless, it would be a valuable research program to challenge this belief by investigating whether our framing and postulates can be extended to other groups, such as non-abelian or non-compact ones.

Looking ahead, our first hope is, again, to steer the debate around QRFs to foundational issues. We deeply believe that getting the concepts right will lay a solid basis for future progress in this field. More specifically, a first direction in which this result could be helpful in the study of compositionality in the field of QRFs. It has been showed that apparent paradoxes appear in such contexts \cite{Angelo_2011}, and it remains an open problem to understand this issue in detail.

Other directions include how to correctly implement perspective changes in the weak approach. Should our argument hold in more general settings, what would be the consequences of the physical relevance of the total charge for these maps? A natural research program would be to understand how to coherently formulate change-of-perspective maps taking into account this new degree of freedom: the total charge. Armed with such new perspective change maps, it would then be relevant to check whether results obtained using other versions of the map, such as the invariance of various informational quantities, would still hold in this context.
\\
\\

\textbf{Note.} In the final stages of this paper, we became aware of independent work by Anne-Catherine de la Hamette, Viktoria Kabel and \v{C}aslav Brukner, who extend the perspectival and perspective-neutral QRF approaches to arbitrary fixed charge sectors and analyse how the accessibility of observables affects which global properties can be inferred from internal perspectives \cite{delaHamette:2025hpd}.

\section*{Acknowledgements}
It is a pleasure to thank Titouan Carette, Esteban Castro-Ruiz, Anne-Catherine de la Hamette, and Thomas Galley for helpful discussions and comments.

The authors are supported by the STeP2 grant (ANR-22-EXES-0013) of Agence Nationale de la Recherche (ANR), the PEPR integrated project EPiQ (ANR-22-PETQ-0007) as part of Plan France 2030, the ANR grant TaQC (ANR-22-CE47-0012), and the ID \#62312 grant from the John Templeton Foundation, as part of the \href{https://www.templeton.org/grant/the-quantum-information-structure-of-spacetime-qiss-second-phase}{‘The Quantum Information Structure of Spacetime’ Project (QISS)}.

\bibliography{refs.bib}{}
\bibliographystyle{utphys}

\appendix

\section{Identifying the mathematical objects embodying the symmetry} \label{ann:sym}
In this Appendix, we expose equivalent ways of describing a symmetry acting on a quantum system with in mind the goal to clarify any ambiguity these multiple language choices can generate. All three can be found in the literature, so that it is important to pin down their relationship.  We insist that this is an independent problem to the one presented in Section \ref{sec:w&s}. Describing the action of a symmetry is different to discussing what it means to be symmetric. In this Appendix, we focus on the former, while Section \ref{sec:w&s} deals with the latter.

When considering a quantum system, there are three equivalent ways to describe how a given symmetry can act on it. One can consider the symmetry-generating observable, the unitary action it is associated to, or the representation of the symmetry group acting on the system. In this section, we restrict to finite Abelian group to keep the discussion centered on the concepts, leaving the extension to groups with less structure to future work.

First, a symmetry can be generated by an observable $O$, which is the physical quantity to which we associate the symmetry. As an example, consider the momentum for the translation symmetry. 
 
The symmetry can also be described by its unitary action. Following up on the translation symmetry example by taking the discrete translation group $G=\Z_n$, the action of the symmetry is given by the unitary $\mathsf{SHIFT}$ operator, which acts on the canonical basis as $\mathsf{SHIFT}\ket{m} = \ket{m+ 1 \mod{n}}, m \in \Z_n$. 

Finally, one can describe the action of symmetry through a unitary representation of its associated group $\mathcal{G}$. This provides us with a collection of symmetry transformations instead of a single operator. In the case of the symmetry under translations in one spatial dimension, one has $\cg=\RR$, yielding uncountably many translation transformations $\hat T(x)$, corresponding to a translation by $x$. 

In order to clarify discussions in the main text, our first point in the Appendix is to show that theses three descriptions are equivalent. The observable is linked to a unitary action through exponentiation $U = e^{iO}$. Conveniently, $U$ and $O$ share the same eigenspaces. This ensures the consistency of our discussion in Section \ref{sec:th} of the main text, when we link the eigenspaces to important mathematical structures in our argument.\footnote{This excludes pathological cases where a couple of eigenvalues of $O$ verify $\lambda_i = \lambda_j +2k\pi, k \in \Z$, but these can be dealt with pretty easily via simply considering a scaled version of the observable $O$. This preserves the eigenspaces, which we are mainly interested in.} Again, in the case of translations, $U$ yields the translated version of a position eigenstate by one unit. 

\begin{figure}
    \centering
    \includegraphics[width=0.9\linewidth]{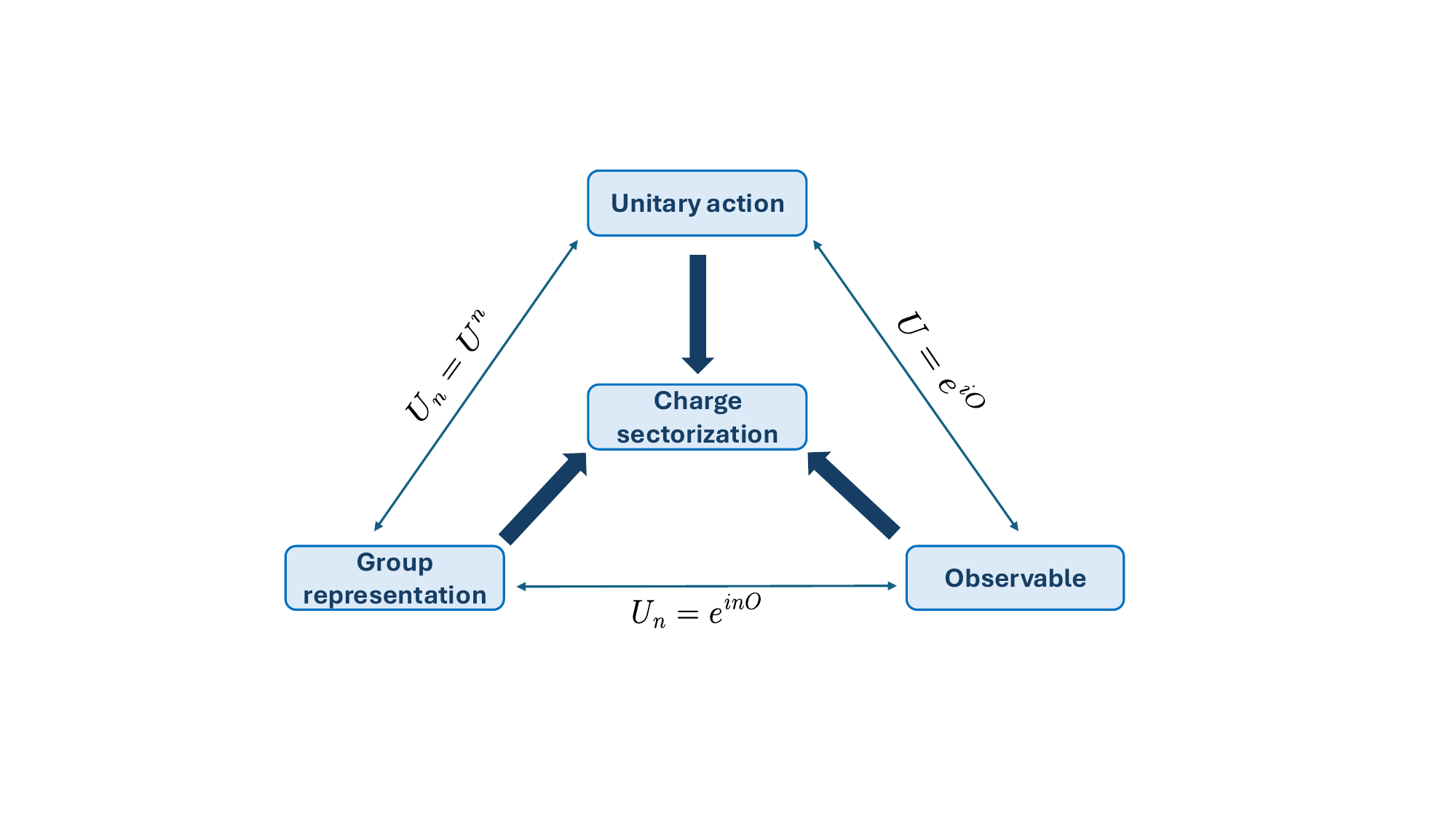 }
    \caption{\em The trinity of approaches to describe a symmetry based on a Abelian cyclic groups}
    \label{fig:tri}
\end{figure}

In the case of finite cyclic Abelian groups, the unitary action and the full unitary are linked through:
\be
U_n = U^n
\ee
where $U_n$ is the representation element corresponding to the $n^{th}$ power of the group generator, and $U$ the one corresponding to the generator, thus also being the aforementioned unitary action of the symmetry,\footnote{For non-cyclic groups, the link is not this straightforward. The key point is the unitary action and the group representation are linked through the sectorization they imply, i.e.\ they are co-diagonalizable. } making it an atomic symmetry transform, that generates all others. For example, for discrete translations, the unitary action generating the symmetry is just a shift of one site, denoted $\mathsf{SHIFT}$, and then the translation by n sites is $\hat T(n) = \mathsf{SHIFT}^n$.

These three approaches give rise to a sectorization of the Hilbert space of the system one constrains, as illustrated in Figure \ref{fig:tri}. As previously stated, describing a symmetry with an observable or the unitary operator corresponding to it via exponentiation yields the same sectorization, which is the eigen-decomposition associated to these operators. The description in terms of unitary group representations also gives rise to a sectorization. Indeed, the unitary operators in the image of the representation of a finite Abelian group respect the group structure, and thus all commute pairwise. This implies that they are all co-diagonalizable, and thus give rise to a sectorization of the Hilbert space they act on. Equivalently, any reducible representation gives rise to block diagonal operators, yielding the same sectorization. 

All three approaches seem to favor one of the spaces (or sectors) in the sectorization they give rise to. If the symmetry observable $O$ has a zero eigenvalue, then it annihilates one sector of the Hilbert space, and in turn, the unitary it is associated to leaves this sector invariant. The ``favored algebra'' is then the algebra of operators that are fully supported on this sector. In the case of group representations, the preferred sector is the one associated with the trivial representation, which also leaves invariant its associated sector. This is the sector the strong approaches singles out as the ``physical'' one.

\section{Translating an argument from \cite{Hoehn:2023ehz} into our language}\label{ann:transl}
\setcounter{equation}{0}
\numberwithin{equation}{section}

We present an argument proposed by Höhn, Kotecha and Mele \cite{Hoehn:2023ehz}, and show how it reduces to the problem of charge accessibility. They explicitly show that two different states in $\ca_{\text{W}}$ are mapped to the same state by the ``gauge-fixing'' (i.e.\ relativization) they use. Let us take a closer look at the argument, and reformulate it in the language we use in the present paper.
Consider $\cg = (\Z_n, + \equiv + \mod n), n < \infty$, and 3 physical systems with state space $\ch = \ch_1 \otimes \ch_2 \otimes \ch_3$. Again, we look at the positions of these systems on a symmetric $n$ site lattice, which are described by states of the form $\ket{i,j,k} \in \ch, i,j,k \in \Z_n$, where the first index is the position of the first system, and so on. The situation is similar to the one described in Example \ref{ex:2}, so the group acts on $\ch$ via a unitary representation $U_g, ~g\in \Z_n$.  Consider now two symmetry-equivalent states:
\be
\ket{\psi} = \ket{0,0, h},~~ \ket{\psi'} = U_g^{\otimes3} \ket{\psi} = \ket{g,g,g+h},~g\neq0
\ee
which we can use to build the two density matrices:
\be
\begin{split}
    \rho &= (a\ket{\psi} + b \ket{\psi'})(a^*\bra{\psi} + b^* \bra{\psi'}) \\
    \rho' &= |a|^2 \ketbra{\psi}{\psi} + |b|^2 \ketbra{\psi'}{\psi'}.
\end{split}
\ee
with $a,b \neq 0$ and $|a|^2+|b|^2 = 1$. Then, the authors compute the image of $\rho$ and $\rho'$ under the action of the weak twirl, to find:
\be
\ct_W(\rho) \neq \ct_W(\rho')
\ee
Then, applying their map of choice to map these states to $\ca_{23|1}$, they find that they get mapped to the same state. The authors then claim this shows that the image of the weak twirl $\ct_W$, i.e.\ $\ca_{\text{W}}$, is too ``large'' to allow for reversible mapping from $\ca_{\text{sym}}$ to the perspectival algebras, and we need to further reduce the allowed state, i.e.\ use $\ca_{\text{sym}}=\ca_{\text{S}}$. In particular they point out how $\ct_W$ allows to distinguish \textit{``states that should only be distinguishable with the help of the external frame''}. What is meant here is that it should not be possible for agents restricting to a single system's perspective (among the three) to tell apart $\rho$ and $\rho'$ since they are a superposition and a mixture of \textit{symmetry equivalent} states.

There are two comments we would like to make about this argument. First, we think it hurts the comprehension of the formalism to view the twirls as operators one can apply to states, as is done here. As we have been trying to make clear, one should rather require invariance under those. We think it can be confusing to talk about $\ct(\sigma) \in \ca_{\text{sym}}$ as the ``invariant part'' of $\sigma \in \ca_{\text{kin}}$, since talking about $\sigma$ in the first place makes no physical sense. Second, we think the way we present the formalism in this paper can help make the argument about the ``external distinguishability'' clearer. To see how, let us dive a bit deeper into the computation of the projections onto $\ca_{\text{W}}$. Using (\ref{eq:tw}), we have 
\be
\begin{split}
\ct_W(\rho) &= \frac{1}{n} \sum_{k=0}^{n-1} U_k \rho U^{\dagger}_k  \\
 &= \frac{|a|^2}{n} \sum_{k=0}^{n-1} U_k \ketbra{\psi}{\psi} U^{\dagger}_k + \frac{ab^*}{n} \sum_{k=0}^{n-1} U_k \ketbra{\psi}{\psi'} U^{\dagger}_k\\
 &+\frac{a^*b}{n} \sum_{k=0}^{n-1} U_k \ketbra{\psi'}{\psi} U^{\dagger}_k + \frac{|b|^2}{n} \sum_{k=0}^{n-1} U_k \ketbra{\psi}{\psi} U^{\dagger}_k \\
 &= \frac{|a|^2}{n} \sum_{k=0}^{n-1}  \ketbra{k,k,k+h}{k,k,k+h} \\
 &+ \frac{ab^*}{n} \sum_{k=0}^{n-1}  \ketbra{k,k,k+h}{g+k,g+k,g+k+h}  \\
 &+\frac{a^*b}{n} \sum_{k=0}^{n-1}  \ketbra{g+k,g+k,g+k+h}{k,k,k+h}  \\
 &+ \frac{|b|^2}{n} \sum_{k=0}^{n-1}  \ketbra{g+k,g+k,g+k+h}{g+k,g+k,g+k+h}. \\
\end{split}
\ee
Now, we extend (\ref{eq:es}) from Example \ref{ex:2} to find the charge eigenstates in the general $n$ case 
\be
\ket{c; r_1, r_2} = \frac{1}{\sqrt{n}} \sum_{k=0}^{n-1} e^{-\frac{2 i \pi}{n} c k} \ket{k, k+r_1, k+r_2}, 
\ee
which we can reverse to get 
\be
\ket{k, k+r_1, k+r_2} =\frac{1}{\sqrt{n}} \sum_{c=0}^{n-1} e^{\frac{2 i \pi}{n} c k} \ket{c; r_1, r_2}.
\ee
Plugging this into the above equation, we get
\be
\begin{split}
    \ct_W(\rho) &= \frac{|a|^2}{n} \sum_{c=0}^{n-1}  \ketbra{c; 0, h}{c;0, h} + \frac{ab^*}{n} \sum_{c=0}^{n-1} e^{-\frac{2 i \pi}{n}c g} \ketbra{c; 0, h}{c;0, h} \\
    &+ \frac{a^*b}{n} \sum_{c=0}^{n-1} e^{\frac{2 i \pi}{n}c g} \ketbra{c; 0, h}{c;0, h} +  \frac{|b|^2}{n} \sum_{c=0}^{n-1}  \ketbra{c; 0, h}{c;0, h} \\
    & = \frac{1}{n} \sum_{c=0}^{n-1} (1 +   ab^*e^{-\frac{2 i \pi}{n}c g} + a^*b e^{\frac{2 i \pi}{n}c g}) \ketbra{c; 0, h}{c;0, h}.
\end{split}
\ee
For $\rho'$, we get the same without the interference terms
\be
\begin{split}
    \ct_W(\rho')
    & = \frac{1}{n} \sum_{c=0}^{n-1} \ketbra{c; 0, h}{c;0, h}.
\end{split}
\ee
We now understand better what should be understood by ``external distinguishability''. The states $\ct_W(\rho)$ and $\ct_W(\rho')$ admit different decompositions with respect to the charge sectorization, i.e.\ they carry different charge information. The authors assume that this information is not internally accessible, which is not compatible with choosing $\ca_{\text{sym}} =\ca_{\text{W}}$ since it contains this information, thus pushing them to choose  $\ca_{\text{sym}} =\ca_{\text{S}}$.

\section{Proof of Theorem \ref{main point}}\label{ann:mainpoint}
\setcounter{equation}{0}
\numberwithin{equation}{section}

Let us take a look at the position basis. As a reminder, this corresponds to taking the factorization $\ch_{\text{\text{kin}}} = \ch_{A|E} \otimes \ch_{B|E}$ in Eve's description, and simply $\ch_{B|A}$ in Alice's view. As we previously stated, in this basis, a beam splitter is simply a Hadamard gate for Alice. Therefore, the diagram corresponding to the measurement of Bob's momentum in Alice's perspective looks like
\be
\begin{quantikz}[wire types = {n, q}]
\lstick{$\ket{0}$} && \targ{} && \rstick{$A$'s local register}\\
& \gate{H} & \ctrl{-1} & \gate{H} &  \rstick{$B|A$} \\
\arrow[from=1-1,to=1-5,black,dashed,-]{}
\end{quantikz}
\ee
In Eve's description, the Hadamard gate is now conjugated with CNOT gates controlled on Alice's position. The writing of the measurement result itself is also conditioned on Alice's position, as we have previously made clear. In Eve's description, the procedure looks like the following:
\be \label{diag2}
\begin{quantikz}[wire types = {n,q,q}]
\lstick{$\ket{0}$} &&&& \targ{} & \targ{} &&&& \rstick{$A$'s local register}\\
& \ctrl{1} &          & \ctrl{1} &          &  \ctrl{-1} & \ctrl{1} &           &\ctrl{1} &\rstick{$A|E$} \\
& \targ{}  & \gate{H} & \targ{}  & \ctrl{-2} && \targ{}  & \gate{H} & \targ{}   &\rstick{$B|E$} \\
\arrow[from=1-1,to=1-10,black,dashed,-]{}
\end{quantikz}
\ee
These two diagrams are the translated version of Figures \ref{fig:A's momentum} and \ref{fig:E_momentum} into the position factorization. One can easily check that the above diagram \ref{diag2} is equal to 

\be \label{diag3}
\begin{quantikz}[wire types = {n,q,q}]
\lstick{$\ket{0}$} &&&& \targ{} &  &&&& \rstick{$A$'s local register}\\
&  &          &  &          &  &             & &\rstick{$A|E$} \\
&   & \gate{H} &   & \ctrl{-2} &&   \gate{H} &   &\rstick{$B|E$} \\
\arrow[from=1-1,to=1-9,black,dashed,-]{}
\end{quantikz}
\ee
In other words, Eve concludes that Alice measured $X_{B|E}$.

This allows us to conclude that the momentum observable included in $\ca_{B|A}$, i.e. the one accessible to Alice, is the same observable as the one accessible to Eve, $X_{B|E} \in \text{Lin}(\ch_{\text{\text{kin}}})$. In a nusthell, one can write $X_{B|E} \in \ca_{B|A}$ and $X_{B|E} = X_{B|A}$.

\end{document}

%% file: commands.tex
\def\be{\begin{equation}}
\def\ee{\end{equation}}
\def\ba{\begin{align}}
\def\ea{\end{align}}

\newcommand{\C}{\ensuremath{\mathbb{C}}}

\newtheorem{theorem}{Theorem}[section]

\newtheorem{definition}{Definition}[section]

\newtheorem{postulate}{Postulate}

\newtheorem{example}{Example}

\DeclareTextFontCommand{\texttt}{\ttfamily\upshape}
\DeclareTextFontCommand{\textrm}{\rmfamily\upshape}

\newcommand{\ca}{\mathcal A}

\newcommand{\cg}{\mathcal G}
\newcommand{\ch}{\mathcal H}

\newcommand{\cs}{\mathcal S}
\newcommand{\ct}{\mathcal T}

\newcommand*{\Z}{\mathbb{Z}}
\newcommand*{\RR}{\mathbb{R}}

\usepackage{stackengine}

\def\yenrule{\rule{1.3ex}{.1ex}}
\def\yen{\renewcommand\stacktype{L}\stackon[.4ex]{\stackon[.65ex]{Y}{\yenrule}}{\yenrule}}

%% file: diagrammes/sample.tikzstyles
\tikzstyle{new style 0}=[fill=white, draw=black, shape=rectangle, minimum width=0.5cm, minimum height=0.5cm]
\tikzstyle{new style 1}=[fill=black, draw=black, shape=circle, diameter=0.1cm]
\tikzstyle{new style 2}=[fill=white, draw=black, shape=beam]

\tikzstyle{dashed}=[-, dashed, dashed pattern=on 4mm off 2mm]